\documentclass[sigconf]{acmart} 
\usepackage{graphicx}
\usepackage{subcaption}
\usepackage{soul}

\AtBeginDocument{%
  }

\usepackage{xcolor,colortbl}

\definecolor{gray}{rgb}{223,223,223}

\usepackage{xspace}
\usepackage{etoolbox}
\usepackage{cleveref}
\usepackage{enumitem}

\newcommand{\system}{RAVEN}

\newcommand{\eg}{\emph{e.g.,\ }}

\newcommand{\etal}{\emph{et al.}\xspace}

\copyrightyear{2026}
\acmYear{2026}
\setcopyright{cc}
\setcctype{by}
\acmConference[CHI '26]{Proceedings of the 2026 CHI Conference on Human Factors in Computing Systems}{April 13--17, 2026}{Barcelona, Spain}
\acmBooktitle{Proceedings of the 2026 CHI Conference on Human Factors in Computing Systems (CHI '26), April 13--17, 2026, Barcelona, Spain}
\acmPrice{}
\acmDOI{10.1145/3772318.3791616}
\acmISBN{979-8-4007-2278-3/2026/04}

\makeatletter
\AtBeginDocument{
\bibcite{PLXBIB0002}{{2}{2024}{{Alharbi et~al{.}}}{{}}}

}
\makeatother

\begin{document}

\title{RAVEN: Realtime Accessibility in Virtual ENvironments for Blind and Low-Vision People}

\author{Xinyun Cao}
\email{xinyunc@umich.edu}
\orcid{0000-0001-9067-062X}
\affiliation{%
  \institution{University of Michigan}
  \city{Ann Arbor}
  \state{Michigan}
  \country{USA}
}

\author{Kexin Phyllis Ju}
\email{kexinju@umich.edu}
\orcid{0009-0002-8272-6552}
\affiliation{%
  \institution{University of Michigan}
  \city{Ann Arbor}
  \state{Michigan}
  \country{USA}
}

\author{Chenglin Li}
\email{lchengl@umich.edu}
\orcid{0009-0008-4335-8885}
\affiliation{%
  \institution{University of Michigan}
  \city{Ann Arbor}
  \state{Michigan}
  \country{USA}
}

\author{Venkatesh Potluri}
\email{potluriv@umich.edu}
\orcid{0000-0002-5027-8831}
\affiliation{%
  \institution{University of Michigan}
  \city{Ann Arbor}
  \state{Michigan}
  \country{USA}
}

\author{Dhruv Jain}
\email{profdj@umich.edu}
\orcid{0000-0001-6176-968X}
\affiliation{%
  \institution{University of Michigan}
  \city{Ann Arbor}
  \state{Michigan}
  \country{USA}
}

\definecolor{alizarin}{RGB}{231, 76, 60}

\begin{abstract}


As virtual 3D environments become more prevalent, equitable access is essential for blind and low-vision (BLV) users, who face challenges with spatial awareness, navigation, and interaction. Prior work has explored supplementing visual information with auditory or haptic modalities, but these methods are static and offer limited support for dynamic, in-context adaptation. Recent advances in generative AI allow users to query and modify 3D scenes via natural language, introducing a paradigm that offers greater flexibility and control for accessibility. We present \system{}, a system that enables BLV users to issue queries and modification prompts to improve the runtime accessibility of 3D virtual scenes. We evaluated \system{} with eight BLV people {and six Unity developers}, generating empirical insights into how conversational programming can support personalized accessibility in 3D environments. Our work highlights both the promise of natural language interaction—intuitive, flexible, and empowering—and the challenges of ensuring reliability, transparency, and trust in generative AI–driven accessibility systems.


\end{abstract}



\begin{teaserfigure}
  \includegraphics[
    width=\textwidth,
    alt={System workflow showing a BLV user modifying a 3D scene. On the left, a low-vision user types the request "Make the sign with the text much bigger" while viewing a game scene. In the center, the system processes the scene objects and user prompt. On the right, the updated scene shows a larger sign, and a text-to-speech message confirms the change.}
  ]{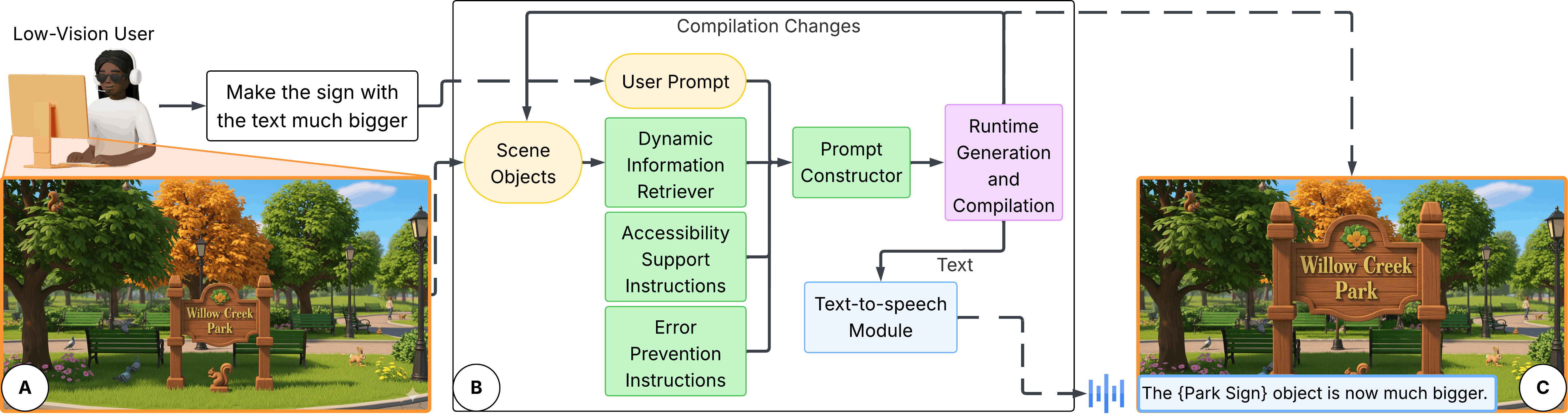}
  \caption{System workflow of \system{}, an interactive tool that enables BLV users to adapt 3D scenes through conversational natural language. The figure illustrates an example interaction: (A) A low-vision user types, \emph{`Make the sign with the text much bigger.''} (B) The system combines semantic scene data with a runtime code-generation LLM to translate this request into an accessibility-enhancing modification. (C) The generated code is compiled and executed in real time, while the system also provides a spoken response confirming the change.}
  \label{fig:teaser}
\end{teaserfigure}
\maketitle


\section{Introduction}
\label{sec:introduction}
Virtual 3D environments have become pervasive, supporting a diverse range of applications such as interactive gaming, social networking, and education. These immersive spaces enable rich interactions, offering users a sense of presence and spatial exploration previously inaccessible in traditional media \cite{lim_gaming_2006, bowman_information-rich_2003}. However, with their increased adoption arises the imperative of ensuring inclusive and equitable access, particularly for blind and low-vision (BLV) users who encounter challenges in spatial understanding, navigation, and object interaction in these environments \cite{white_toward_2008}.
s
To address these challenges, prior work proposed tools that modify or supplement visual information using alternative modalities, such as audio descriptions \cite{gonzalez-mora_seeing_2006}, haptic feedback \cite{jansson_haptic_1999, zhao_enabling_2018}, and enhanced visual effects tailored for low vision \cite{zhao_seeingvr_2019}. Systems like SceneWeaver \cite{balasubramanian_enable_2023} offer users greater agency over when and how to consume scene descriptions. Commercial platforms have also adopted accessibility features—such as high-contrast display modes and spatial audio—to accommodate BLV players in mainstream video games \cite{bayliss_fortnite_2022, playstation_last_2024}. These methods allow BLV users to utilize pre-defined tools to enhance the accessibility of a virtual 3D environment.

However, current approaches to accessibility share a fundamental limitation: they are predominantly developer-driven and static, such as fixed mappings for color changes or auditory overlays, and may not align with the nuanced and evolving needs of individual users \cite{creed_inclusive_2024}. Moreover, they often require users to learn specific control mappings or adjust settings in non-intuitive ways, leading to a steep learning curve while offering limited support for dynamic, context-specific adaptation. 

Recent advances in generative AI (GenAI), particularly large language models (LLMs), open possibilities for more flexible and conversational interaction. LLMs have been increasingly applied to accessibility tasks such as querying images \cite{adnin_i_2024} and runtime scene editing \cite{jennings_whats_2024}, suggesting the potential for users to directly query and adapt 3D scenes through natural language, bypassing rigid developer-defined workflows and expanding user agency.

Motivated by this potential, we present \system{}, an interactive system that empowers BLV users to engage with 3D scenes via natural language interaction. \system{} supports both querying (\eg{}``What’s around me?'') and accessibility-related modifications (\eg{}``Make the table brighter'' or ``Move the bench closer''). It integrates LLMs with semantic scene data and runtime code generation to apply changes at runtime, while providing spoken responses to user queries. Interaction is iterative, enabling users to refine modifications through follow-up prompts in a dialogue-like flow.

We evaluated \system{} with eight BLV participants in a user study including three interactive scenarios: a guided tutorial, structured tasks, and free exploration. {The scenarios and study tasks were designed around six accessibility categories drawn from prior work.} Participants engaged in conversational interactions to tailor scenes according to their accessibility preferences, offering quantitative feedback on usability and qualitative reflections on their experiences. Findings reveal critical insights into the potential and limitations of generative AI-supported interactions, highlighting both opportunities for enhanced accessibility and challenges with trust and reliability. {We also conducted a preliminary study with six Unity developers to evaluate the system’s usability from a developer perspective, yielding insights into its learnability, overall usability, and opportunities for improvement.}

In summary, our work contributes: (1) the design and implementation of \system{}, a GenAI-powered system for real-time, natural language-driven querying and modification of 3D environments for accessibility, and (2) empirical insights from a study with eight BLV users {and a preliminary study with six Unity developers}, highlighting interaction strategies, usability, and challenges that inform future LLM-based accessibility tools. Our work demonstrates the importance of flexible, intuitive, and contextually adaptive accessibility in virtual environments.  



\section{Related Work}
\label{sec:related-work}
We review prior work on accessibility in virtual 3D environments for BLV people, survey how generative AI has been used to enhance access across contexts, and examine runtime generative tools for virtual 3D scenes that inform our system’s capabilities.

\subsection{BLV Accessibility in Virtual 3D Environments}
\label{ssec:a11y-in-v3d}

{For the context of this work, virtual 3D environments are computer-generated three-dimensional spaces that may be displayed on PCs, mobile devices, or immersive VR headsets.} Early efforts created bespoke virtual environments for BLV users \cite{sanchez_virtual_1999, gonzalez-mora_seeing_2006, picinali_exploration_2014, andrade_echohouse_2018}, for  activities including rehabilitation and orientation training programs~\cite{seki_training_2011, tzovaras_interactive_2009} and  specific experiences such as boxing~\cite{furtado_designing_2025}, racing~\cite{gluck_racing_2021}, and table tennis~\cite{kamath_playing_2024}. While these systems demonstrate feasibility and value, their design assumptions often limit generalizability beyond the particular contexts for which they were built.

To broaden access in general-purpose 3D spaces,  researchers have explored substituting visual information via audio and haptics. Guerreiro \etal{} articulate a design space for auditory substitution in virtual environments, providing a theoretical framework for cue design~\cite{Guerreiro_design_2023}. Commercial haptic hardware, however, typically lacks the spatial resolution to convey rich scene structure, so haptics is commonly paired with audio~\cite{Nair_navstick_2021, balasubramanian_enable_2023} or requires custom devices~\cite{zhao_enabling_2018, tzovaras_interactive_2009, yuan_blind_2008}. A growing thread emphasizes agency and free exploration. Canetroller provides an auditory-haptic “white cane,” enabling transfer of real-world cane skills into VR for learnable exploration~\cite{zhao_enabling_2018}. NavStick supports surveying surroundings through a gamepad thumbstick with auditory feedback, improving mental-map accuracy over menu-based baselines~\cite{Nair_navstick_2021}. SceneWeaver gives users control over when and how to receive descriptions, increasing exploratory agency~\cite{balasubramanian_enable_2023}. This body of work shows how auditory and haptic feedback can help BLV users understand and navigate 3D scenes.

For users with residual vision, enhancing the visual modality itself can be effective. SeeingVR demonstrates how magnification, contrast enhancement, recoloring, and related tools help low-vision users complete tasks more quickly and accurately~\cite{zhao_seeingvr_2019}. 

Many of these ideas have influenced commercial games, which now include accessibility settings such as navigation assistance, combat audio cues, and high-contrast modes~\cite{bayliss_fortnite_2022,playstation_last_2024}, offering evidence of long-term learnability and adoption.

Despite these advances, most approaches remain developer-defined and static: substitutions (\eg{}auditory cues) and modifications (\eg{}color mappings) are predetermined and may not track the diverse, evolving needs of BLV users~\cite{creed_inclusive_2024}. They can also impose learning burdens through mode switches, keybindings, or complex settings, which is especially challenging for newcomers~\cite{Kreimeier_two_2020, maidenbaum_perception_2016}. To our knowledge, \system{} is the first system to enable BLV users to \emph{query} and \emph{modify} virtual 3D scenes \emph{at runtime} via natural language, shifting control from developer-defined presets to user-directed, in-situ adaptation.

Despite these advances, most approaches remain developer-defined and static: substitutions (\eg{}auditory cues) and  modifications (\eg{}color mappings) are predetermined and may not track the diverse, evolving needs of BLV users~\cite{creed_inclusive_2024}.  They can also impose learning burdens through mode switches, keybindings, or complex settings, which is especially challenging for newcomers~\cite{Kreimeier_two_2020, maidenbaum_perception_2016}.

{We also build on a prior non-archival ASSETS demo that presented an early prototype based on our work ~\cite{cao_demo_2025}. That demo introduced the idea of using LLM-driven runtime modifications in Unity and reported preliminary results, but it did not include the full analysis of the BLV user study or the developer study reported here. In this paper, we substantially extend that prototype by (1) fully specifying the system architecture and prompt-engineering framework (\cref{sssec:SceneConstruction}-\cref{sssec:sys-summary}), (2) deriving and documenting accessibility rules and categories from BLV literature, and (3) contributing two new empirical evaluations: a three-scene study with eight BLV participants and a preliminary usability study with six Unity developers. We therefore treat the demo as an early proof-of-concept and this paper as the first archival presentation of the complete system and its evaluation.}

To our knowledge, \system{} is the first system to enable BLV users to \emph{query} and \emph{modify} virtual 3D scenes \emph{at runtime} via natural language, shifting control from developer-defined presets to user-directed, in-situ adaptation.

\subsection{Generative AI for Visual Accessibility}
\label{ssec:genAI-4-a11y}

Generative AI (GenAI) tools have increasingly supported BLV users’ everyday access needs, including real-world scene understanding~\cite{chang_worldscribe_2024, bemyeyes_introducing_2025}, navigating digital interfaces~\cite{kodandaram_enabling_2024}, visual authoring~\cite{chang_editscribe_2024}, and professional productivity~\cite{perera_sky_2025, seo_maidr_2024}. These systems leverage natural language as an intuitive control channel and adapt responses to context and user goals. For example, Savant allows users to control screen readers conversationally, reducing dependence on complex shortcut vocabularies and lowering workload~\cite{kodandaram_enabling_2024}. WorldScribe provides real-time, context-aware descriptions, offering concise updates for dynamic scenes and more detailed accounts for stable ones~\cite{chang_worldscribe_2024}. This line of work motivates us to introduce similarly flexible, conversational interaction to virtual 3D contexts.

GenAI also enables visual content creation, spanning images~\cite{zhao_advances_2024, florian_introduction_2024} and 3D scenes~\cite{hu_scenecraft_2024, zhang_vrcopilot_2024}, opening avenues for creative agency among BLV users~\cite{bennett_painting_2024}. Yet generated content is often inaccessible to its creators, complicating evaluation, verification, and iterative refinement~\cite{huh_GenAssist_2023}. To address this in 2D settings, GenAssist combines vision-language and language models to pose verification and style questions, and synthesize question-guided descriptions that help users assess prompt alignment and compare candidates~\cite{huh_GenAssist_2023}. AltCanvas introduces a spatially structured, tile-based interface to improve layout comprehension while adding, editing, and moving visual components~\cite{lee_altcanvas_2024}. {While effective in 2D, analogous accessibility support for \emph{virtual 3D} generative workflows remains underexplored, where BLV users face additional demands for spatial and embodied understanding.}

Adoption of GenAI in accessibility also introduces risks. A recent study by Glazko \etal{} \cite{glazko_autoethnographic_2023} found that LLMs could recite accessibility guidance but struggled to apply it in context~\cite{glazko_autoethnographic_2023}; an empirical audit found high rates of accessibility issues in generated web code, with many defects persisting even after remediation prompts~\cite{aljedaani_does_2024}. Despite these errors, the apparent fluency of agents can foster false expectations, overconfidence, and inaccurate mental models~\cite{glazko_autoethnographic_2023, adnin_i_2024}. The potential for hallucination further necessitates trustworthy verification and avenues for user contestation~\cite{alharbi_misfitting_2024,adnin_i_2024,zhang2025llmhallucinationspracticalcode}. Finally, training data can encode biases and ableist stereotypes~\cite{glazko_autoethnographic_2023,urbina_disability_2025, gadiraju_disability_2023, glazko_identifying_2024, mack_wheelchair_2024}. \system{} attempts to address these concerns by embedding specific accessibility instructions into prompts for the Unity3D environment, lowering sampling temperature to reduce randomness~\cite{vellum_llm_2025}, and explicitly detecting “out-of-scope” user requests to avoid spurious outputs. 

Overall, \system{} leverages LLM-generated responses and code to support accessibility in virtual 3D environments, aiming to combine the flexibility of conversational interfaces with safeguards that mitigate known GenAI limitations.

\subsection{Runtime Generation in Virtual 3D Scenes Using GenAI}
\label{genAI-4-v3d}

Recent work demonstrates GenAI systems for 3D scene \emph{creation}~\cite{hu_scenecraft_2024, Yang_2024_CVPR, zhang_vrcopilot_2024}, \emph{modification}~\cite{jennings_whats_2024, delatorre_llmr_2024, chen_analyzing_2025}, and \emph{interaction}~\cite{liang_handproxy_2025, hu_gesprompt_2025}. Collectively, these systems show that models can reason about spatial layouts, process multi-modal inputs, and generate executable changes-capabilities that are promising for accessibility.

Creation-focused systems primarily support 3D designers. SceneCraft generates Blender-executable Python scripts from language, rendering assets programmatically~\cite{hu_scenecraft_2024}. HoloDeck uses GPT-4 to select and place objects from model collections to instantiate embodied AI environments~\cite{Yang_2024_CVPR}. VRCopilot supports multi-modal furniture layout design to enhance immersion and creativity~\cite{zhang_vrcopilot_2024}. These systems illustrate how GenAI can interpret spatial intent and produce code to realize 3D content.

Beyond creation, GenAI can edit scenes and generate behaviors at runtime. GROMIT encodes Unity scenes as semantic graphs to enable LLM-authored behaviors and was evaluated with developers creating new game mechanics~\cite{jennings_whats_2024}. LLMR provides a framework for real-time creation and modification of mixed-reality experiences~\cite{delatorre_llmr_2024}. Chen \etal{} analyze user input patterns for LLM-assisted manipulation in immersive settings, identifying design considerations around user agency and handling uncertainty and hallucination~\cite{chen_analyzing_2025}. This line of work suggests rich opportunities for accessibility-oriented modifications.

Other systems employ LLMs to mediate interaction in 3D spaces. HandProxy lets users command a virtual hand via natural language to operate UI and manipulate objects~\cite{liang_handproxy_2025}. GesPrompt augments this paradigm with co-speech gesture for lower-effort, more intuitive control~\cite{hu_gesprompt_2025}. These approaches echo non-GenAI proxies such as VoiceAttack~\cite{voiceattack_voiceattack_2025} but also leverage multi-modal interaction for greater expressiveness and ease of use.

Despite these advances, utilizing runtime generation for \emph{accessibility} remains underexplored. While systems such as LLMR hint at adaptations for color blindness, near-sightedness, or child-friendly design~\cite{delatorre_llmr_2024}, they do not address the breadth of BLV needs nor evaluate with BLV users. Our work is the first to investigate LLM-driven, runtime scene querying and modification for accessibility with and for the BLV community.



\section{Iterative System Design}
\begin{figure*}
    \centering
    \includegraphics[
    width=1\linewidth,
    alt={A collage of images showing video game scenes. One image is in the center, pointing arrows towards six images that surround it. The texts on the arrows are commands: Make the car bright yellow; Move me next to the policeman; Make the car plate bigger; Make the sunlight brighter; Make the pedestrian louder; Increase policeman's pitch. The six surrounding images show the corresponding changes.}
    ]{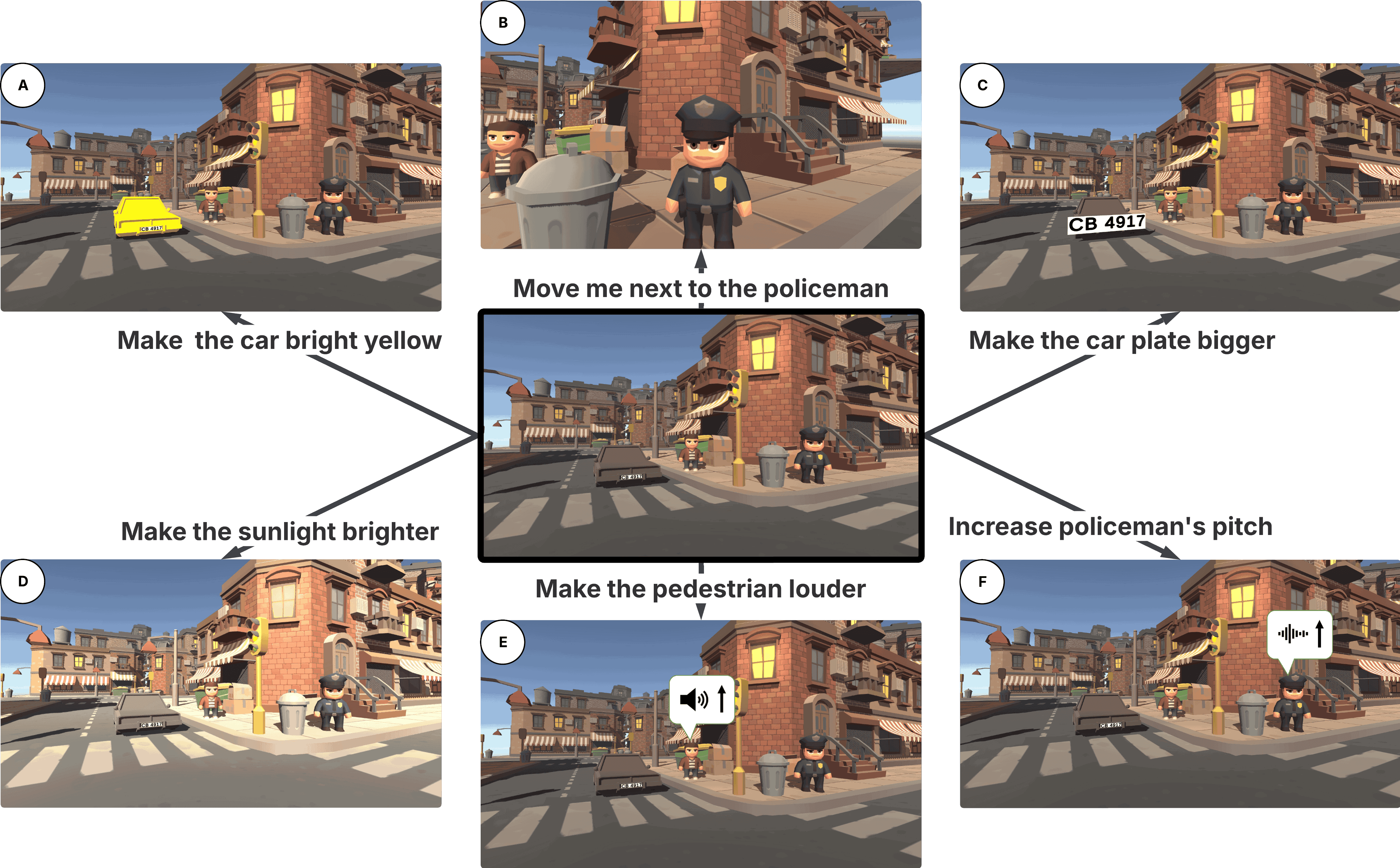}
    \caption{Examples of system usage. The center image shows the original game scene, with six surrounding panels illustrating accessibility-driven modifications: (A) change object color, (B) reposition the player, (C) enlarge a text object, (D) increase brightness, (E) amplify audio volume, and (F) adjust audio pitch. Bubbles and icons in E and F visualize auditory changes. These are only a few of the many types of modifications \system{} can support to enable flexible, user-driven accessibility.}
    \label{fig:system-usage-examples}
\end{figure*}

\label{sec:systemDesign}
\system{} supports real-time queries and modifications in virtual 3D environments to improve accessibility for BLV users. {The system is designed to be open, where users can use any natural language prompt. However, to ground its utility in user needs, we first developed an initial prototype and conducted a pilot study with three BLV participants.} Insights from this study informed iterative refinements, resulting in the final system presented here. In this section, we introduce an illustrative use scenario (\cref{ssec:Scenario}), describe the pilot study and resulting design modifications (\cref{ssec:Pilot}), and detail the final system architecture (\cref{ssec:System}).

\subsection{Use Scenario}
\label{ssec:Scenario}
Consider Dez, a low-vision gamer who sometimes uses screen readers but primarily relies on residual vision augmented with accessibility modifications. While playing a 3D game enhanced by \system{}, Dez first requests a description of the scene. Within seconds, the system replies: \emph{``You are located at a dark street corner\ldots In front of you, there is a mysterious parked car\ldots Further in the distance, two characters are talking on their phones\ldots''}

Dez then asks the system to brighten the scene. In response, the system adds point light sources, which Dez further refines by requesting they be placed near the mysterious car. Curious, he follows up: \emph{``How many lights are near the car now?''}, and the system provides a count. After iteratively adjusting brightness levels, Dez continues customizing the environment—highlighting the car in bright yellow, raising the pitch and volume of a key character’s dialogue, and enlarging the license plate text.

This scenario illustrates how users can iteratively adapt scene elements through natural language, tailoring the environment to their accessibility needs. \Cref{fig:system-usage-examples} highlights additional examples, showing that \system{} can support a wide range of modifications.

\subsection{Pilot Study and Iterative Modifications}
\label{ssec:Pilot}
We built an initial prototype of \system{} based on the intended use scenario and insights from prior BLV accessibility research~\cite{mukherjee_semantic_2004, Nair_navstick_2021, white_toward_2008}. The prototype included keyboard shortcuts for selecting and bookmarking objects as well as a natural language interface for querying and modifying the environment. To evaluate its feasibility and identify areas for refinement, we conducted a pilot study with three BLV participants (two with residual vision and one blind).  Participants explored two scenes designed to present distinct accessibility challenges: (1) a static scene with complex spatial relationships to probe exploration, and (2) a noisy scene with overlapping conversations to simulate dynamic social interactions. After receiving training in system interactions, participants freely explored the scenes. We collected usability ratings, observational data, and semi-structured interview feedback, which informed the iterative modifications described below.

\subsubsection{Removing Keyboard Shortcuts for Conversation-Only Interaction}
\label{ssec:pilot-finding-learning-curve}

In the initial prototype, participants could interact with the system in two ways: through keyboard shortcuts (\eg{}selecting and bookmarking objects) and through natural language commands for querying and modification. Usability ratings indicated a mixed experience, with an average System Usability Scale (SUS) score of 75.83—suggesting reasonable usability but leaving room for improvement. Participants reported that remembering keyboard shortcuts created “\emph{initial frustrations}” and described them as non-intuitive. Consistently, the selection and bookmarking features tied to keyboard commands received the lowest ratings. In contrast, natural language interaction was rated most highly and described as more intuitive. To reduce learning burden and cognitive load, we removed keyboard-based features and focused exclusively on conversational interaction.

\subsubsection{Adding Egocentric Spatial Descriptions}
\label{ssec:pilot-finding-directions}

In the initial prototype, the system described object locations using raw 3D coordinates (\emph{x, y, z}). Participants reported that this format was unintuitive and difficult to map onto their own perspective in the scene. For example, one participant remarked, “\emph{I don’t quite understand the coordinates},” while another noted that the system failed to recognize egocentric references such as “\emph{what’s in front of me}.” To address this limitation, we augmented object metadata with embodied spatial relations (\emph{in front of}, \emph{to the left/right}, \emph{behind}) and relative distance from the player. We then instructed the LLM to use this metadata to generate egocentric, perspective-aware descriptions that aligned with how participants naturally referenced space.

\subsubsection{Mitigating Errors and Hallucination}
\label{ssec:pilot-finding-hallucination}

During the pilot, participants observed typical LLM shortcomings, including hallucination, randomness, and failure to handle vague requests gracefully. At times the system claimed to have executed a modification when none had occurred, or produced erroneous changes. To mitigate these issues, we added prompt-engineering strategies instructing the system to: (1) ask clarifying questions when requests are vague, (2) acknowledge and recover from errors reported by users, and (3) recognize out-of-scope requests (\eg{}adding magnifiers) rather than attempting unsupported modifications. Although hallucination remains a limitation of generative AI technologies~\cite{openai2024gpt4technicalreport}, these strategies improved the system’s reliability and transparency.

\subsubsection{Summary of Iterative Refinements}
\label{ssec:mod-summary}

Through iteration, we streamlined \system{} toward conversational interaction, refined spatial reasoning with embodied relations, and improved robustness against hallucination. Participants navigate scenes using standard keyboard controls (arrow keys for movement; W,A,S,D for camera panning), while all accessibility queries and modifications are handled through natural language. This design reduces learning overhead while preserving the flexibility and power of real-time modifications.

\subsection{The \system{} System}
\label{ssec:System}

The final \system{} architecture (\cref{fig:teaser}-B) requires only minimal developer annotation during scene creation (\cref{sssec:SceneConstruction}) and integrates several components to support natural language interaction. At runtime, the system self-voices user input (\cref{sssec:TTS}), retrieves an accessibility-augmented semantic scene graph (SSG) from the environment (\cref{sssec:DynamicInformationRetrieval}), and constructs prompts that combine user queries with accessibility and error-prevention instructions (\cref{sssec:PromptConstructor}). These prompts are executed through the GROMIT runtime generation system~\cite{jennings_whats_2024}, which generates Unity code to implement requested modifications and returns a textual response. This response is announced to the user through text-to-speech, completing the interaction loop. {The full system implementation is open-sourced and can be accessed here: \url{https://github.com/SoundabilityLab/RAVEN}.}

\subsubsection{Scene Construction}
\label{sssec:SceneConstruction}

To prepare a scene, developers create a 3D environment in Unity following standard workflows, then identify a subset of important objects ({\eg{}}game items on a table). For each selected object, {developers attach a lightweight Unity script and indicate whether it is physical ({\eg{}}a table) or non-physical ({\eg{}}an ambient sound source), and whether it is the player. Optional developer-provided descriptions of visual or auditory properties may also be included.} Developers also include lightweight scripts that connect the scene to the LLM agent. This metadata is the only additional input required, keeping developer burden minimal.

\subsubsection{Text-to-Speech for Self-Voicing}
\label{sssec:TTS}

Standard screen readers do not integrate well with Unity environments, yet audio feedback is critical for BLV users. To address this, \system{} is designed as a self-voicing application: it announces characters as they are typed, reads words upon completion, and vocalizes system responses. We implemented this feature by building a wrapper around the Microsoft Speech API for Unity~\cite{microsoftTTS}.

\subsubsection{Dynamic Information Retriever}
\label{sssec:DynamicInformationRetrieval}

The Dynamic Information Retriever produces an up-to-date semantic scene graph (SSG) {each time a prompt is issued}, ensuring that the LLM receives current environmental context. Building on the structure introduced by Jennings \etal{}~\cite{jennings_whats_2024}, we extend the SSG with accessibility metadata. For each relevant object, the SSG stores its name, developer-provided description (visual, auditory, and functional properties), attached scripts, position, scale, and hierarchical relationships. {To support accessibility queries and modifications, the following metadata is recomputed at every prompt:}

\begin{enumerate}
    \item Color (HEX code).
    \item Text content and font size.
    \item Egocentric direction (\emph{in front of}, \emph{to the left/right}, \emph{behind}) and distance from the player.
    \item Light source density.
    \item Audio source status (mute/unmute, volume, pitch, and range).
\end{enumerate}

This enriched SSG provides the LLM with the contextual grounding necessary for accurate descriptions and modifications.

\subsubsection{Prompt Constructor}
\label{sssec:PromptConstructor}

The Prompt Constructor fuses user input with scene data and prompt-engineered instructions before sending requests to the LLM. Prior research shows that LLMs often struggle with accessibility alignment and error handling~\cite{glazko_autoethnographic_2023, ji_ai_alignment_2025, shen_bidirectional_2024}. To address this, the Prompt Constructor integrates two sets of instructions, which are included with every request as part of the conversational history.

\textbf{1. Accessibility Support Instructions.}  
Although LLMs {may} possess accessibility knowledge, they may struggle to apply it effectively in specific contexts~\cite{glazko_autoethnographic_2023}. {We therefore synthesized accessibility needs in virtual 3D environments from prior BLV accessibility research~\cite{white_toward_2008, Nair_navstick_2021, Guerreiro_design_2023, zhao_seeingvr_2019}. These needs span color/contrast adjustments, brightness changes, spatial understanding, text enlargement, scene description, and audio manipulation. We operationalized these needs into a set of explicit rules (\eg{}how to adjust Unity C\# variables, how to simplify materials before recoloring, and how to convert Unity terminology into user-facing language) that guide the LLM’s descriptions and modifications.}

\textbf{2. Error Prevention Instructions.}  
To mitigate hallucinations and improve transparency, we designed instructions that enable the system to handle ambiguity and errors conversationally. If a request is incomplete or vague, the LLM asks clarifying questions (\eg{}\emph{``It seems like your request is not clear. Could you please provide more details or clarify what you would like to achieve?''}). If a user reports an error (\eg{}\emph{``it's not working''}), the system is instructed to acknowledge the limitation and suggest alternative approaches. Finally, the LLM is primed~\cite{sun2025newdatapermeatesllm} with a list of out-of-scope tasks identified during the pilot study (\eg{}adding magnifiers), ensuring that unsupported requests are clearly surfaced rather than producing erroneous code.

Together, these measures ground the LLM in accessibility context, reduce hallucinations, and preserve the flexibility of conversational interaction. The full prompt is provided in \cref{ssec:appendix1}.

\subsubsection{Runtime Modification Generation and Compilation}
\label{sssec:Gromit}

Once the prompt is constructed, it is sent to GROMIT~\cite{nicholasjj_nicholasjjgromit_2024, jennings_whats_2024}, an open-source runtime behavior generation system for Unity. GROMIT processes the prompt by generating Unity code that implements the requested modification, attaches the compiled script to the relevant object, and returns a textual response. This enables the system to dynamically update scenes while providing users with immediate, self-voiced feedback. {We used GPT-4o for language and code generation, as it was the most advanced model that supported feasible real-time integration during system development.}

\subsubsection{System Design Summary}
\label{sssec:sys-summary}

In sum, \system{} introduces a new approach to making virtual 3D environments accessible: BLV users can issue natural language queries and modifications that are executed in real time. This is enabled by three key contributions: (1) a self-voicing interface that supports prompt entry and feedback, (2) an accessibility-augmented semantic scene graph that encodes perspective-aware relations and multimodal attributes, and (3) prompt-engineering strategies that align the LLM with accessibility goals and mitigate hallucinations. Together, these components shift control from static, developer-defined accommodations to dynamic, user-directed accessibility.



\section{User Study Method}
\label{sec:method}

To evaluate the utility of on-demand accessibility modifications and the usability of \system{}, we conducted a user study with BLV participants. The study focused on six accessibility categories (\cref{ssec:A11YModCategories}) and three scenarios with progressively open-ended tasks (\cref{ssec:scene-description}). We addressed two research questions:

\begin{itemize}
    \item \textbf{RQ1:} Can \system{} support blind and low-vision people in experiencing a 3D scene?
    \item \textbf{RQ2:} What kinds of prompts do BLV participants use when interacting with the system to improve accessibility?
\end{itemize}

\subsection{Accessibility Categories}
\label{ssec:A11YModCategories}

{To scaffold our study design, we synthesized six accessibility categories from prior work across game accessibility toolkits and empirical studies of BLV user needs~\cite{szpiro_how_2016,wurm_color_1993,rigden_eye_1999,uysa_writing_2012,rubin_the_2006,zhao_seeingvr_2019}. These categories were used solely to structure our evaluation tasks and analysis; they do not constrain \system's capabilities nor restrict what users may request during interaction.} Four categories address the visual domain and two target the auditory dimension.

\subsubsection{Visual Categories}
These categories allow users to adapt the visual presentation of a scene:
\begin{enumerate}
    \item \emph{Color:} Retrieve or modify color schemes and object colors to aid recognition, especially for color-blind and low-vision users~\cite{wurm_color_1993,rigden_eye_1999}.
    \item \emph{Object Location:} Retrieve or reposition objects (\eg{}furniture, characters) to simplify navigation and interaction, inspired by assistive game toolkits~\cite{voiceattack_voiceattack_2025}.
    \item \emph{Object Size:} Resize objects or text to enhance visibility and readability~\cite{uysa_writing_2012,rubin_the_2006}.
    \item \emph{Scene Brightness:} Query or adjust overall brightness or specific light sources to accommodate visual sensitivity~\cite{zhao_seeingvr_2019}.
\end{enumerate}

\subsubsection{Auditory Categories}
These categories adapt the auditory experience of a scene:
\begin{enumerate}
    \item \emph{Audio Volume:} Increase or decrease the volume of sound sources to isolate or emphasize elements~\cite{chang_soundshift_2024,rychtarikova2015blind}.
    \item \emph{Audio Pitch:} Alter pitch to distinguish sounds, highlight important content, or emulate screen reader cues~\cite{chang_soundshift_2024,grussenmeyer_accessible_2017,voiceover,nvda}.
\end{enumerate}

\subsection{Scenes}
\label{ssec:scene-description}

We designed three scenes to scaffold participants’ exploration of the six accessibility categories, with increasing complexity and freedom of interaction (\cref{fig:scenes}).

\subsubsection{Scene 1: Guided Tutorial}
The first scene introduced participants to the system in a simple, game-like room containing two geometric objects with distinct colors, two speakers playing dialogue, a torch with continuous sound and animation, a text object, and a table (\cref{fig:scenes}a). The researcher demonstrated each category using a three-step prompt pattern (ask, modify, verify), after which participants practiced similar prompts themselves. This structure, inspired by prior work on generative AI accessibility tools~\cite{glazko_autoethnographic_2023,adnin_i_2024,huh_GenAssist_2023}, ensured that participants could explore both querying and modification. Full list of demo prompts used are included in~\cref{ssec:appendix2}.

\subsubsection{Scene 2: Task-Driven Exploration}
\label{sssec:scene2}
The second scene, a cat park with natural elements, three cats producing distinct meows, a bench, a streetlamp, and a garden hut (\cref{fig:scenes}b), was paired with six predefined tasks (\cref{tab:scene2-task}). The tasks were designed to align with the six accessibility categories. After completing a task, participants reflected on their experience, including whether the interaction was intuitive, useful, or confusing. This setup allowed us to observe how participants applied categories in a goal-oriented context.

\begin{table*}
  \caption{Tasks in Scene 2}
  \label{tab:scene2-task}
  \begin{tabular}  {c p{13.5cm}}
    \toprule
    Number & Task\\
    \midrule
    \texttt 1 & Ask the system to describe the scene.\\
	     \cellcolor[HTML]{dfdfdf}{2} & \cellcolor[HTML]{dfdfdf}{The bench is a color unfriendly to low-vision users, can you change it to a better color?}\\
    \texttt 3 & Notice that there are some cats in the scene making some sounds. What’s different about how each of the cats sounds? Which of these cats seems to be the happiest?\\
     \cellcolor[HTML]{dfdfdf}{4} & \cellcolor[HTML]{dfdfdf}{Benches in the park are sometimes dedicated to people who donated money for them, and they have their name written in a short text on the benches. Where is the bench and who is this bench dedicated to?}\\
    \texttt 5 & Imagine you’re taking a (virtual) photo of the white cat, try increasing the brightness of the scene for a better photo.\\
 \cellcolor[HTML]{dfdfdf}{6} & \cellcolor[HTML]{dfdfdf}{Make the bench bigger so that you and the cat have space to sit together.}\\
    \bottomrule
  \end{tabular}
\end{table*}

\subsubsection{Scene 3: Open-Ended Exploration}
\label{sssec:scene3}
\begin{figure}[t]
    \centering
    \begin{subfigure}[t]{0.75\linewidth}
        \centering
        \includegraphics[
        width=1\textwidth,
        alt={An image of a video game scene. There is a white table with a green sphere, a red cube and a burning torch on top of it. Some text is hovering over the table.}
        ]{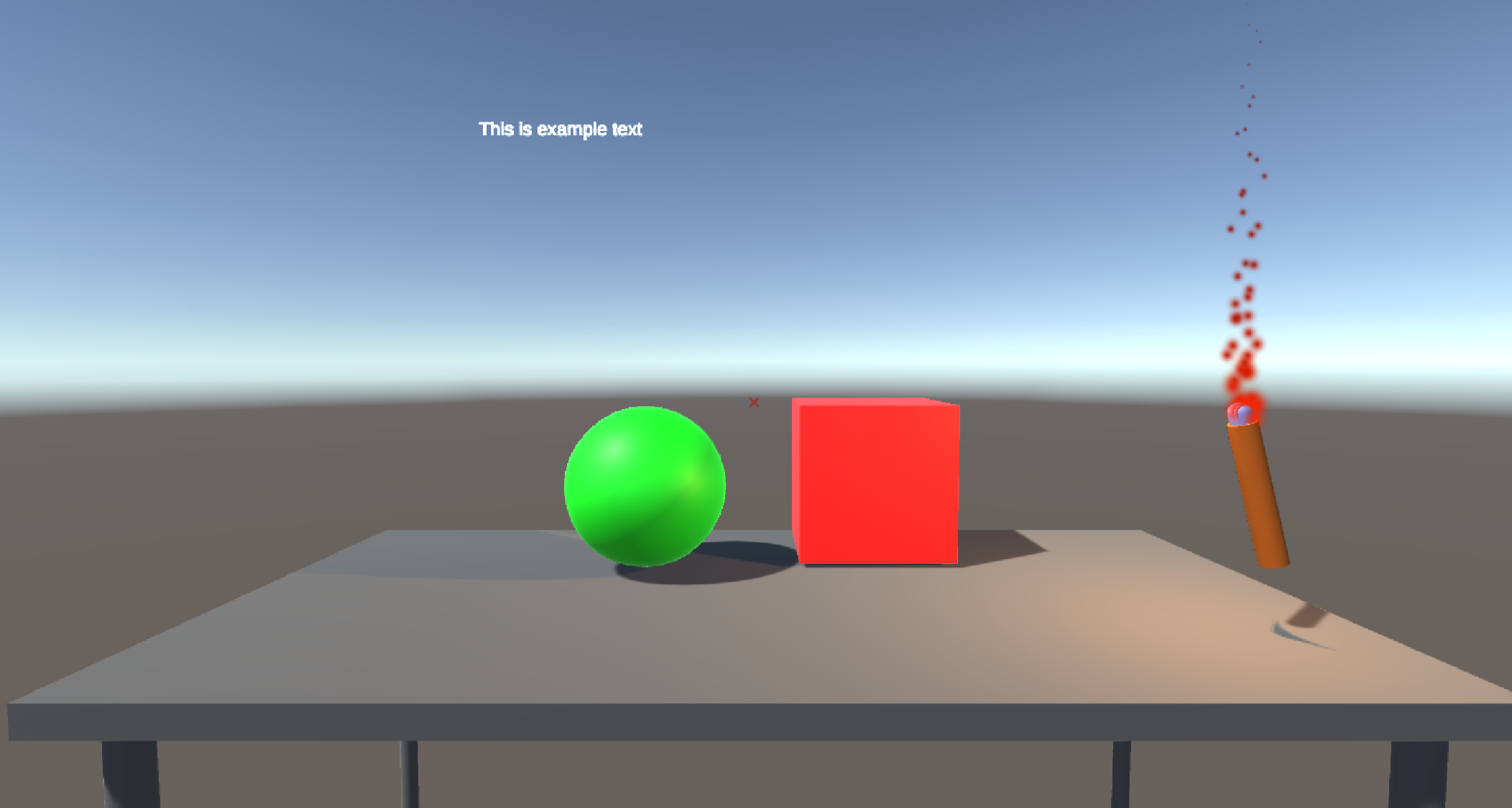}
        \caption{Scene 1}
        \label{fig:sceneDemo}
    \end{subfigure}
    \begin{subfigure}[t]{0.75\linewidth}
        \centering
        \includegraphics[
        width=1\textwidth,
        alt={An image of a video game scene in a park. There is a road extending from left to right into the distance. A bench under a street lamp is next to the road. Several cats are on the road. There's a garden hut in the distance.}
        ]{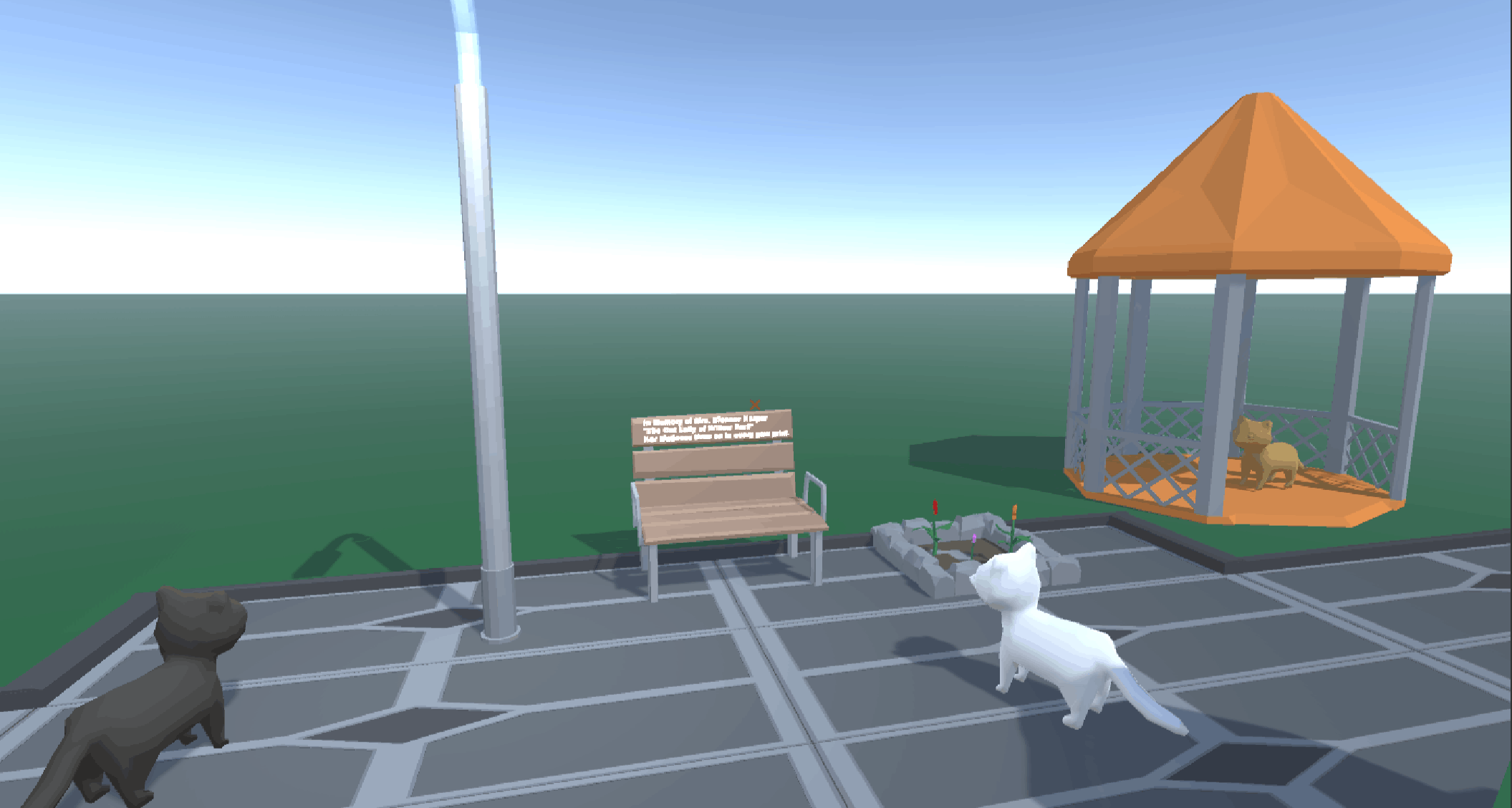}
        \caption{Scene 2}
        \label{fig:sceneCatPark}
    \end{subfigure}
    \begin{subfigure}[t]{0.75\linewidth}
        \centering
        \includegraphics[
        width=1\textwidth,
        alt={An image of an indoor video game scene. There are two tables in the room. The table on the right has colorful vases and small items on top. There's a piece of white text on the wall above this table. The table on the left is surrounded by colorful chairs and has small items on top.}
        ]{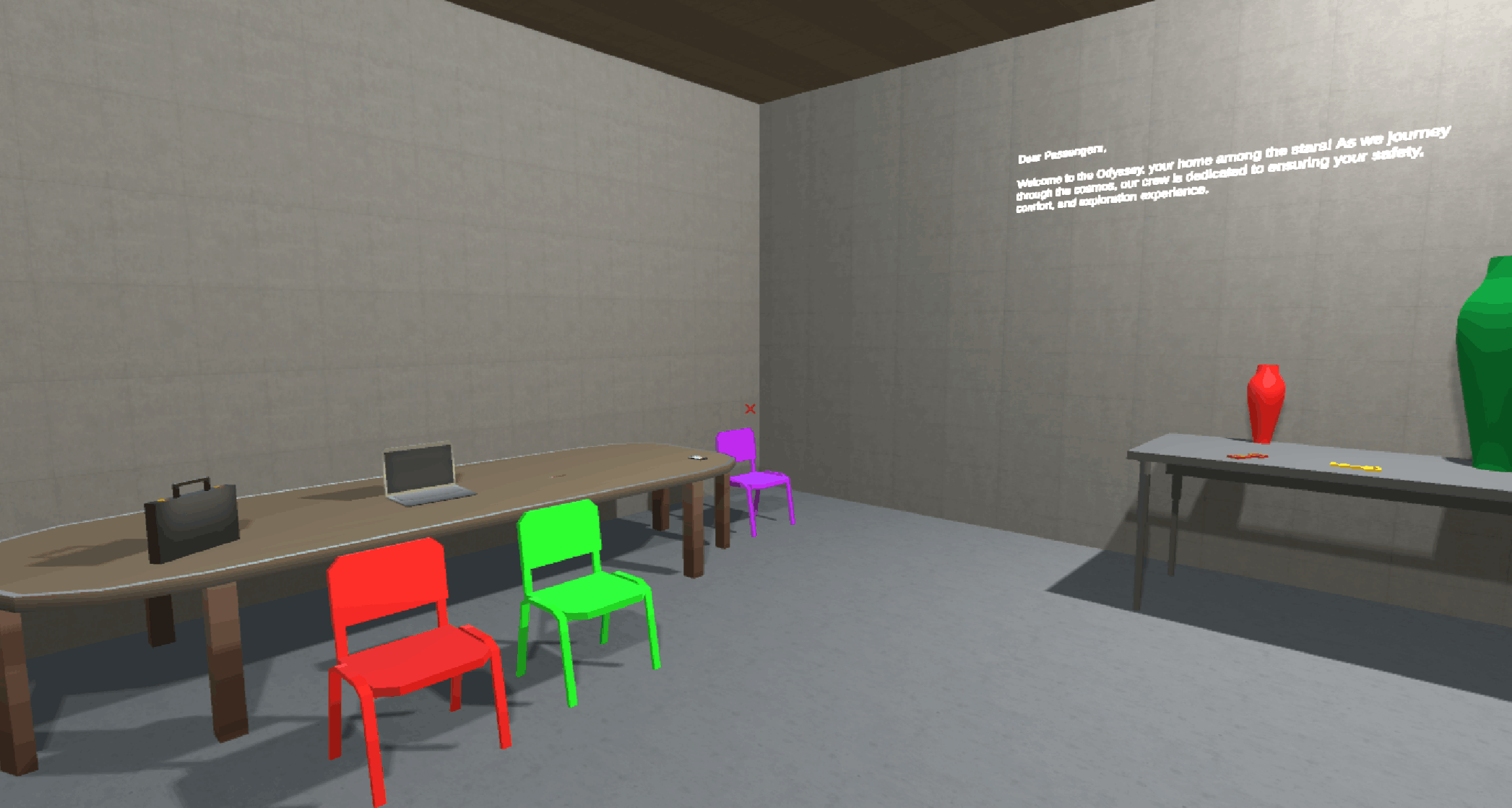}
        \caption{Scene 3}
        \label{fig:sceneRoom}
    \end{subfigure}
    \caption{Screenshots of the three scenes used in our evaluation. Scene 1: a demo with simple objects and sound sources. Scene 2: a park with cats meowing and background nature sounds. Scene 3: a spaceship room with furniture, small items, and sci-fi sound sources.}
    \label{fig:scenes}
\end{figure}

The final scene encouraged open-ended use of the system. Participants explored a spaceship-themed room containing sixteen objects and three sound sources (\cref{fig:scenes}c). They were given ten minutes to query, modify, and adapt the environment according to their own needs, without predefined tasks. This scenario revealed how participants appropriated the system in less structured contexts.

\subsection{Study Design}
\label{ssec:study_design}

\subsubsection{Participants}
\label{sssec:participants}
We recruited eight BLV participants (five men, three women; ages 24–50, $M$=36.6, $SD$=7.9) through mailing lists and word of mouth. Two participants were blind with no vision, two had light perception, and four had low vision. All used screen readers; some also used braille displays, large text, or commercial interpreting services (\eg{}Be My Eyes, Aira). Three regularly used conversational AI apps (\eg{}ChatGPT, Claude), four had limited experience, and one had never used them. Three had played games prior to vision loss, while five continued gaming using accessibility features. Full demographics are shown in \cref{tab:participants}.

\begin{table*}
  \caption{Participant demographics. Visual ability labels: B = blind, LP = blind with light perception, VI = visually impaired (all self-identified). “Onset” indicates the age at which vision loss began. “AT usage” lists accessibility technologies used. “LLM usage” refers to experience with conversational AI apps.}
  \label{tab:participants}
  \setlength{\tabcolsep}{3pt}
  \begin{tabular}{lll p{3.5cm} l p{3cm} l p{3cm}}
    \toprule
    Code & Age & Gender & Visual Ability & Onset & AT usage & LLM usage & Video game usage\\
    \midrule
    \texttt P1-LP & 34 & Man & Light perception only & 26 & Screen readers & Never used & Played a few\\
    \cellcolor[HTML]{dfdfdf}{P2-VI} & \cellcolor[HTML]{dfdfdf}{43} & \cellcolor[HTML]{dfdfdf}{Woman} & \cellcolor[HTML]{dfdfdf}{Blurry and muted vision} & \cellcolor[HTML]{dfdfdf}{Birth} &  \cellcolor[HTML]{dfdfdf}{Screen readers, braille display, and AI apps} & \cellcolor[HTML]{dfdfdf}{Tried a few times} & \cellcolor[HTML]{dfdfdf}{Played a few as kid}\\
    \texttt P3-VI & 40 & Man & Severe tunnel vision & Birth & Screen readers & Used regularly & Played a lot as kid, only haptic games recently\\
    \cellcolor[HTML]{dfdfdf}{P4-LP} & \cellcolor[HTML]{dfdfdf}{41} & \cellcolor[HTML]{dfdfdf}{Man} & \cellcolor[HTML]{dfdfdf}{Blind with light perception} & \cellcolor[HTML]{dfdfdf}{Birth} & \cellcolor[HTML]{dfdfdf}{Screen readers, braille display, and AI apps} & \cellcolor[HTML]{dfdfdf}{Used regularly} & \cellcolor[HTML]{dfdfdf}{Played a few}\\
    \texttt P5-VI & 28 & Man & Astigmatism in right eye & 5 & Large text, screen readers & Tried a few times & Played a few\\
    \cellcolor[HTML]{dfdfdf}{P6-VI} & \cellcolor[HTML]{dfdfdf}{24} & \cellcolor[HTML]{dfdfdf}{Woman} & \cellcolor[HTML]{dfdfdf}{Legally blind with Retinopathy of Prematurity} & \cellcolor[HTML]{dfdfdf}{Birth} &  \cellcolor[HTML]{dfdfdf}{Large text, screen readers} & \cellcolor[HTML]{dfdfdf}{Tried a few times} & \cellcolor[HTML]{dfdfdf}{Played a few as kid}\\
    \texttt P7-B & 33 & Man & Totally blind with no light perception & Birth &  Screen readers, braille display & Tried a few times & Play audio games regularly\\
    \cellcolor[HTML]{dfdfdf}{P8-B} & \cellcolor[HTML]{dfdfdf}{50} & \cellcolor[HTML]{dfdfdf}{Woman} & \cellcolor[HTML]{dfdfdf}{Totally blind with no usable vision} & \cellcolor[HTML]{dfdfdf}{5} &  \cellcolor[HTML]{dfdfdf}{Screen readers, braille display, and AI apps} &\cellcolor[HTML]{dfdfdf}{Used regularly} & \cellcolor[HTML]{dfdfdf}{Never since vision loss}\\
    \bottomrule
  \end{tabular}
\end{table*}

\subsubsection{Study Setup}
\label{sssec:study-setup}
\begin{figure}
    \centering
    \includegraphics[
    width=0.75\linewidth,
    alt={The image shows the study setup. A BLV participant and a researcher are sitting at a table with a laptop. The laptop screen is showing Scene 2 with RAVEN in action. The BLV user is using a keyboard to interact with the scene. Both the participant and the researcher had stereo headphones to access the sounds from the scene.}
    ]{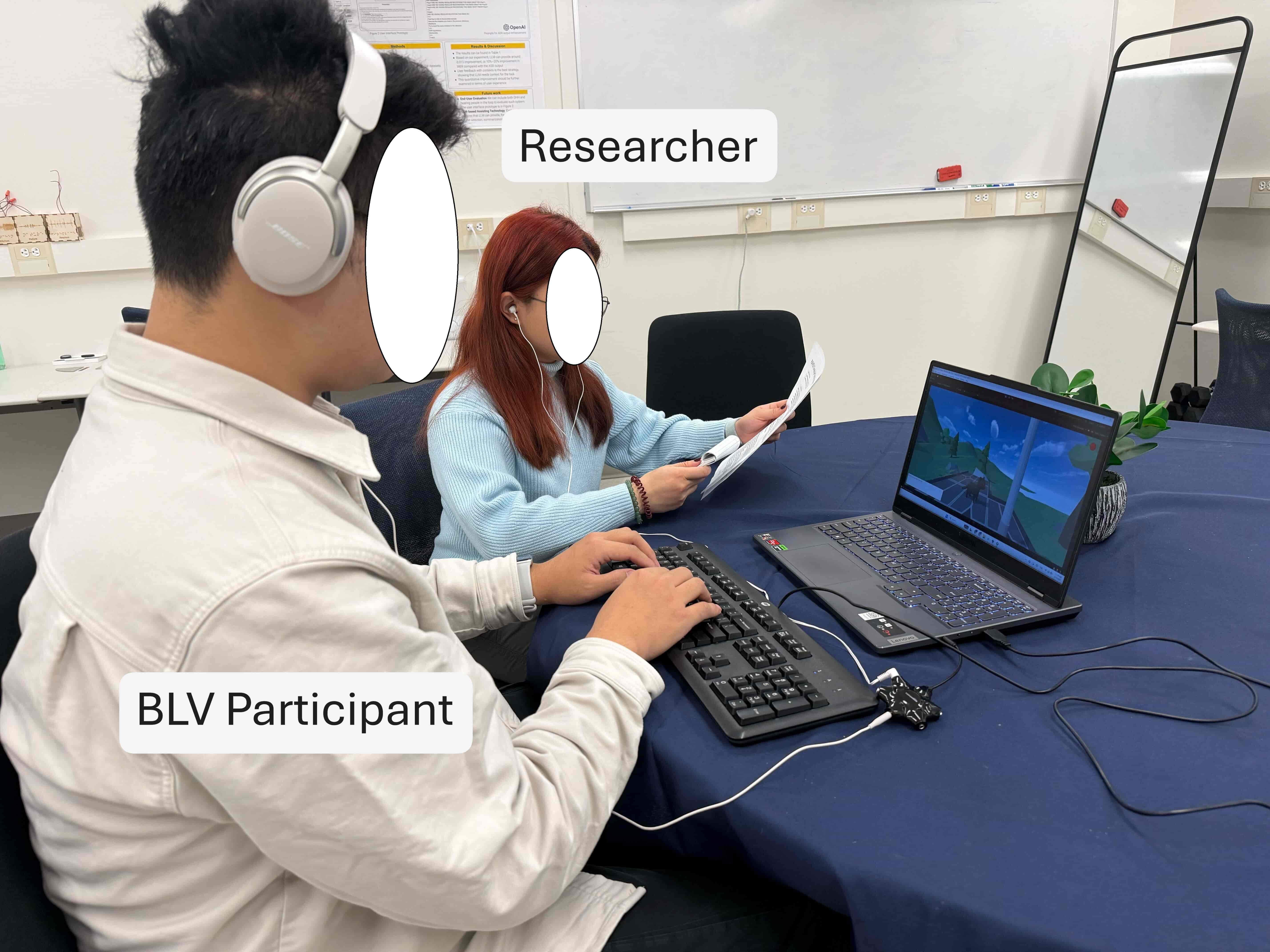}
    \caption{{The study setup for the evaluation of RAVEN with BLV participants.}}
    \label{fig:study-setup-image}
\end{figure}

{The scenes in the study were deployed in a 3D environment running on a laptop. Prior work shows that this setup could be effectively used to evaluate visual accessibility in 3D and 360 environments~\cite{Nair_navstick_2021, Froehlich_streetviewAI_2025}. We excluded more novel formats (\eg{}VR) to keep the study centered on 3D spatial aspect and to avoid confounding the results with extra learning demands. Participants used a laptop device running Unity, a full-sized keyboard, and stereo headphones. The researcher also used stereo headphones to monitor system audio output in real time. See \cref{fig:study-setup-image}.}

\subsubsection{Procedure}
Studies lasted approximately 1.5 hours and were conducted either in-lab (\emph{N=}6) or at participants’ homes (\emph{N=}2). Each session followed the three scene explorations (\cref{ssec:scene-description}), then a survey (usefulness of categories, confidence, intuitiveness, and SUS~\cite{brooke_sus_1996}), and finally a semi-structured interview (usability, new category ideas, prompt strategies, accessibility needs, AI trust). Sessions were audio- and video-recorded. Participants received a \$50 gift card and travel reimbursement. The study was IRB-approved.

\subsubsection{Data Analysis}

We analyzed three sources of data: survey ratings, prompt logs, and interview transcripts. Together, these provided complementary perspectives on participants’ experiences and interaction strategies. {The research team is mixed-ability, with one BLV, one DHH, and three able-bodied researchers. We acknowledge that these backgrounds shape our interpretation.}

\paragraph{Survey Data.}  
{The survey was designed to address RQ1, evaluating \system{}'s support for BLV users in 3D scenes. It included an SUS questionnaire~\cite{brooke_sus_1996} to evaluate system usability, and 5-point Likert scale questions about: usefulness of categories (\cref{ssec:A11YModCategories}), user confidence in accessibility improvement, and intuitiveness of the system.} We computed descriptive statistics (means, standard deviations) to summarize the data.

\paragraph{Prompt Logs.}
Prompts and system responses from Scenes 2 and 3 were recorded, yielding 336 valid prompts (181 from Scene 2 and 155 from Scene 3), after excluding the tutorial scene and erroneous inputs. {We collected response time from recordings, measuring from the frame of user input submission to the frame when the LLM finished loading and reply appeared.} Each prompt was independently coded by multiple researchers along three dimensions, with disagreements resolved through discussion:
\begin{enumerate}
    \item \emph{Correctness:} whether the system response successfully achieved the request (success), correctly identified the request as out-of-scope, misaligned with user intent (intent error), or failed against ground truth (technical error).  
    \item \emph{Command category:} the type of modification or query. Six predefined categories guided the study design, while analysis revealed four emergent categories, producing ten in total. Nine prompts were coded as “Other,” and eight combined multiple categories. We also calculated correctness rates per category.  
    \item \emph{User goal:} the underlying intent behind the prompt (\eg{}creative modification, verification). Through iterative coding, we developed three code groups encompassing eight codes. Prompts could receive multiple goal codes to capture complex intentions.  
\end{enumerate}

\paragraph{Interview Data.}
Interviews were transcribed and analyzed using applied thematic analysis~\cite{guest_applied_2012}. Three researchers collaboratively: (1) familiarized themselves with the transcripts, (2) tagged data with codes, (3) developed an initial codebook, (4) refined codes and themes, (5) finalized the codebook, and (6) synthesized themes into findings. 

This process produced a final codebook with 114 codes, organized into 30 third-level themes, 12 second-level themes, and 4 first-level themes. {Before integrating qualitative insights into the Findings section, we summarize the structure of our thematic analysis. Our final codebook contained four major themes and twelve subthemes: (1) \textbf{System Performance and Reliability} (perceived accuracy and trust, handling of hallucinations, and verification strategies); (2) \textbf{Interaction and Feedback Experience} (intuitiveness of natural language interaction, appropriation of iterative prompting, and modality preferences across audio and visual cues); (3) \textbf{Prompt Categories and Use Patterns} (task-driven use in Scene~2, open-ended exploration in Scene~3, and differences between effective and ineffective prompting strategies); and (4) \textbf{Ethical and Broader Impacts} (concerns about safety and scene integrity, desires for guardrails and transparency, and reflections on broader accessibility opportunities and risks). These themes structure the qualitative findings reported in \cref{sec:study-findings} and clarify how we organized perspectives on system performance, interaction experience, prompting behavior, and broader impacts.}



\section{User Study Findings}
\label{sec:study-findings}
Our analysis highlights both the promise and current limitations of \system{}. We first report {the system's performance}, followed by participants’ overall perceptions of usability, then examine performance across the six \emph{guiding} and four \emph{emerging} prompt categories, and finally describe the strategies and goals that shaped how participants engaged with the system.

\subsection{System Performance Evaluation}
{Of the 336 valid prompts from Scenes~2 and 3 in the user study, 253 (75.3\%) produced a correct query answer or intended modification (determined through researcher coding of session recordings), 9 (2.7\%) were correctly flagged as out-of-scope, and 74 (22.0\%) failed. Failures included \emph{intent errors} (14 prompts, where the LLM misunderstood participant goals) and \emph{technical errors} (60 prompts, where the LLM hallucinated, misrepresented system capabilities, or failed in code execution).}

{The average response time was 3.1 seconds ($SD{=}3.7$) across all prompts. A breakdown of the average response time for different correctness labels is shown in \cref{tab:prompt-error-rate-response-time}. \emph{Acknowledge Out-of-scope} prompts had the shortest average response time and the lowest variability. The remaining correctness labels had similar average response times, with \emph{Technical Errors} requiring slightly longer.}

\begin{table*}
  \caption{{Counts and percentages of prompts per correctness label, and their corresponding average response times.}}
  \label{tab:prompt-error-rate-response-time}
  \begin{tabular}{lllll}
    \toprule
    Correctness Label & Count & Percentage (\%) & Avg. Response Time (sec) & SD of Response Time (sec)\\
    \midrule
    \texttt Success & 253 & 75.3 & 3.1 & 4.1 \\
    \cellcolor[HTML]{dfdfdf}{Acknowledge Out-of-scope} & \cellcolor[HTML]{dfdfdf}{9} & \cellcolor[HTML]{dfdfdf}{2.7} & \cellcolor[HTML]{dfdfdf}{2.1} & \cellcolor[HTML]{dfdfdf}{0.5} \\
    \texttt Intent Errors & 14 & 4.2 & 3.2 & 1.6 \\
    \cellcolor[HTML]{dfdfdf}{Technical Errors} & \cellcolor[HTML]{dfdfdf}{60} & \cellcolor[HTML]{dfdfdf}{17.9} & \cellcolor[HTML]{dfdfdf}{3.4} & \cellcolor[HTML]{dfdfdf}{2.5} \\
    \texttt Total Errors (Intent and Technical) & 74 & 22.0 & 3.2 & 2.3 \\
     \cellcolor[HTML]{dfdfdf}{All Prompts} & \cellcolor[HTML]{dfdfdf}{336} & \cellcolor[HTML]{dfdfdf}{-} & \cellcolor[HTML]{dfdfdf}{3.1} & \cellcolor[HTML]{dfdfdf}{3.7} \\
    \bottomrule
  \end{tabular}
\end{table*}

\subsection{Overall System Usability}
\label{ssec:system-usability}

Participants rated the system positively across confidence, intuitiveness, and usability. On average, they reported confidence in using the tool to make scenes accessible (\(M=4.1, SD=0.8\), scale 1–5, 5 being best) and rated the system as intuitive (\(M=4.3, SD=0.9\)). The mean SUS score was 79.7, corresponding to an A– on the Sauro–Lewis curved grading scale~\cite{sauro_quantifying_2016} and categorized as ``good'' usability~\cite{Bangor_SUS_adjective}.

Subjective feedback echoed these results. P1-LP observed that the tool \emph{``seems very good at allowing someone to get a sense of the sort of overall environment they're in and the sort of properties of items within that environment.''} P5-VI appreciated that adjustments could be made in-scene without interrupting gameplay, and P7-B compared the learnability favorably against audio games that require memorizing shortcuts: \emph{``you could tell the AI to change something that you forgot the keyboard command for and then it could help you.''} Several participants (N=2) highlighted robustness to typos, ambiguity, and compound requests. Others (N=2) praised the open-endedness, with P8-B noting: \emph{``the sky was the limit in some of the things that I could ask [the system] to do.''}  

Five participants felt the system could enhance the gaming experience for BLV players. P6-VI described how it expanded her interest: \emph{``[There were] games that I previously weren't super interested in because of all the barriers related to the visual barriers that I face. [Having this tool] could potentially increase my interest in computer games.''} Participants also envisioned applications beyond games, including web design (P5-VI) and educational software (P4-LP), and for broader groups such as players with cognitive disabilities (P1-LP) or colorblindness (P7-B).

Despite these advantages, several limitations emerged. P4-LP described the gaps in flexible language and cross-modal consistency: extinguishing a torch stopped the light but left the fire sound. P6-VI similarly noted misapplied assumptions, such as turning all objects white after asking to \emph{``make the room white''}. Participants (N=7) raised trust concerns about misleading responses. As P2-VI asked, \emph{``If it gives the wrong information, how is a visually-impaired person going to know?''} P4-LP concluded: \emph{``right now, [the system] is not so foolproof as to say I completely trust it.''} These findings highlight both the promise and current fragility of the approach. {Because our participants’ visual abilities spanned a continuum—from no usable vision to varying forms of low vision—these impressions reflect a spectrum of experiences rather than a clean blind/low-vision split. This motivates our choice to report visual-ability differences qualitatively instead of conducting binary group comparisons.}

\subsection{Performance and Usage Across Prompt Categories}
\label{ssec:prompt-categories-performance}

Qualitative coding of 336 prompts revealed ten categories: six \emph{guiding categories} defined in study design (Object Location, Audio Volume, Color, Object Size, Scene Brightness, Audio Pitch) and four \emph{emerging categories} that surfaced during analysis (Scene Description, Semantic Description, Functionality, Creation/Deletion). Nine prompts were coded as ``Other'' and eight combined multiple categories (\eg{}\emph{``Make the bench normal-sized (Object Size) and put us on the bench'' (Object Location)}). See \cref{tab:categorizationCode} for code definitions and examples.  

\begin{figure}[t]
    \centering
    \begin{subfigure}[t]{0.49\textwidth}
    \vspace{0pt}
        \includegraphics[
        width=\textwidth,
        alt={The image shows a histogram of the total occurrence of each prompt category, sorted from highest to lowest for guiding categories and for emerging categories. For the guiding categories, from the highest to lowest are Object Location being higher than 70; then Color and Audio Volume at around 40; Object Size between 30-40; Scene Brightness at around 30, Audio Pitch at around 10. For emerging categories, from highest to lowest are Scene Description at around 40, Semantic Description at around 30, Functionality at around 10, and Creation and Deletion at around 5.}
        ]{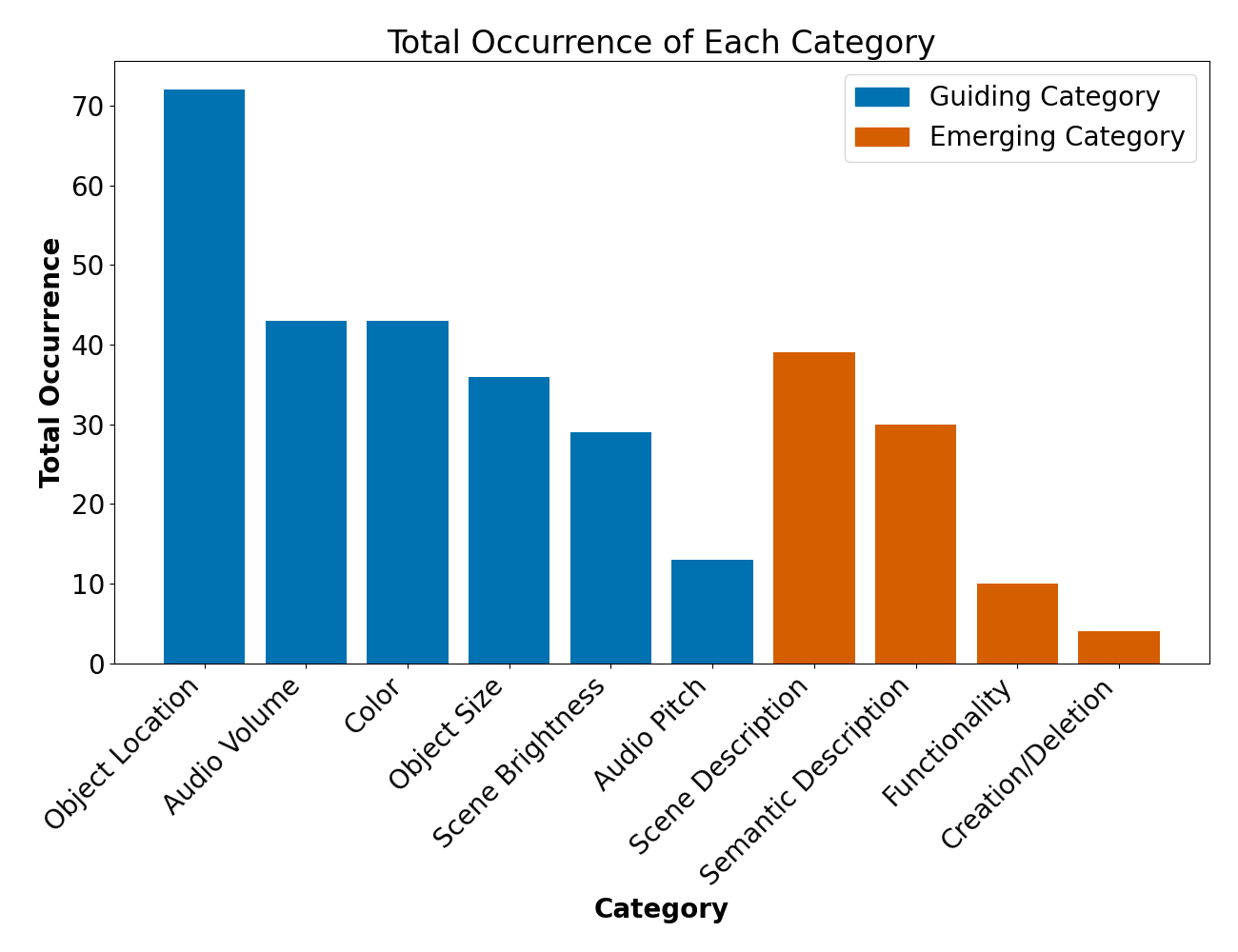}
        \caption{Occurrences of each category (blue = guiding, orange = emerging).}
        \label{fig:category-occurrences}
    \end{subfigure}
    \hfill
    \begin{subfigure}[t]{0.49\textwidth}
    \vspace{0pt}
        \includegraphics[
        width=\textwidth,
        alt={The image shows a stacked bar chart of the percentage of correctness codes breakdown by categories. The last column shows the percentage data for all prompts combined.}
        ]{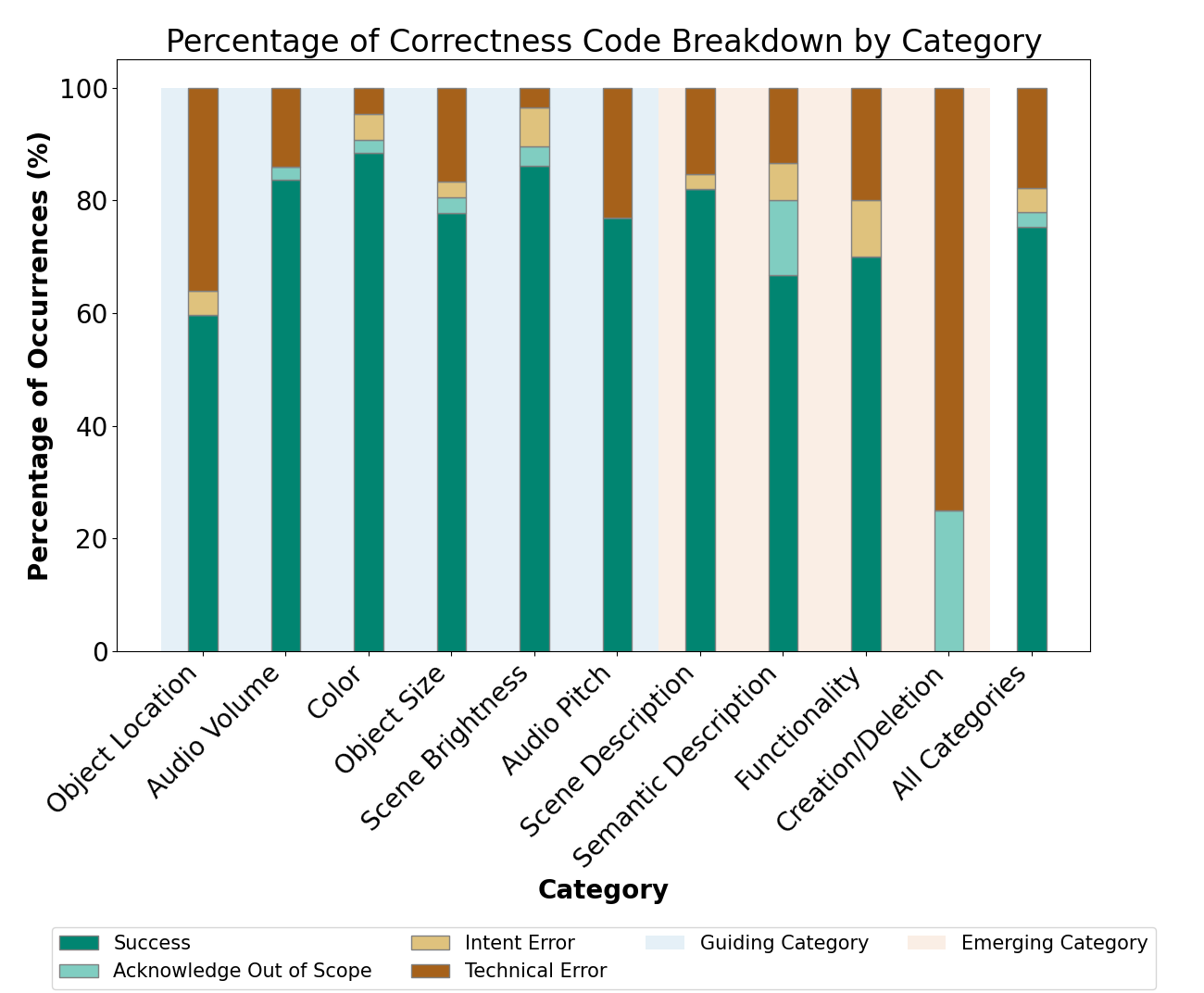}
        \caption{Correctness breakdown per category.}
        \label{fig:category-success}
    \end{subfigure}
    \caption{Prompt usage and correctness across categories.}
    \label{fig:categories}
\end{figure}

\Cref{fig:categories} summarizes category occurrence and correctness. Object Location was most frequent, but with a relatively high error rate. Color and Audio Volume were also common, with high success rates. Semantic Description appeared often but was limited by out-of-scope acknowledgments. Functionality and Creation/Deletion were rare and error-prone, reflecting unsupported user needs.

\begin{table*}
  \caption{Prompt category codes with descriptions and examples.}
  \label{tab:categorizationCode}
  \begin{tabular}{l p{5.5cm} p{5.5cm}}
    \toprule
    Code & Description & Example \\
    \midrule
    \texttt Object Location & Modify or describe the location of object(s) or the player. & ``Move me closer to the white cat.'', ``Where is the gold key?'' \\
    \cellcolor[HTML]{dfdfdf}{Audio Volume} & \cellcolor[HTML]{dfdfdf}{Modify or describe the loudness of sound sources.} & \cellcolor[HTML]{dfdfdf}{``Mute the laptop.'', ``How loud is each cat?''} \\
    \texttt Color & Modify or describe the color of object(s). & ``Make the bench bright yellow.'', ``What color are the keys?'' \\
    \cellcolor[HTML]{dfdfdf}{Object Size} & \cellcolor[HTML]{dfdfdf}{Modify or describe the size of object(s) or text.} & \cellcolor[HTML]{dfdfdf}{``Increase the length of the bench.'', ``How big is the laptop?''} \\
    \texttt Scene Brightness & Modify or describe brightness of lights or the scene. & ``Increase skylight intensity.'', ``How bright are the lamps?'' \\
    \cellcolor[HTML]{dfdfdf}{Audio Pitch} & \cellcolor[HTML]{dfdfdf}{Modify or describe pitch of sound sources.} & \cellcolor[HTML]{dfdfdf}{``Lower pitch of laptop to 0.4.'', ``Which cat has the high-pitched meow?''} \\
    \texttt Scene Description & Describe the general scene or search for object(s). & ``Describe the scene around me.'', ``Does the room have a window?'' \\
    \cellcolor[HTML]{dfdfdf}{Semantic Description} & \cellcolor[HTML]{dfdfdf}{Describe objects based on semantic qualities.} & \cellcolor[HTML]{dfdfdf}{``Which cat seems happiest?'', ``What is on the laptop screen?''} \\
    \texttt Functionality & Use objects functionally. & ``Open the laptop.'', ``Answer the phone.'' \\
    \cellcolor[HTML]{dfdfdf}{Creation/Deletion} & \cellcolor[HTML]{dfdfdf}{Add or remove objects.} & \cellcolor[HTML]{dfdfdf}{``Add a canopy to the bench.'', ``Make the pen disappear.''} \\
    \bottomrule
  \end{tabular}
\end{table*}

\subsubsection{Guiding Categories}

\paragraph{Object Location.}  
Most frequently used (\(M=4.8, SD=0.7\) usefulness), this category enhanced spatial awareness through directional cues and repositioning. P4-LP stressed: \emph{``You have to know how to orient in the world, and you have to be able to have it clear in space.''} Despite high value, success rates were lower than average due to LLM hallucinations (\eg{}misaligned perspectives) and the system’s reliance on object-center coordinates, which provided only coarse control. This worked for broad moves (\eg{}placing the player in a room) but broke down for finer actions (\eg{}putting a phone on a chair without it floating between the legs). Some participants preferred richer directional schemas (clock-face or cardinal points) over simple left/right cues, highlighting opportunities for more precise spatial references.

\paragraph{Audio Volume.}  
Highly rated (\(M=4.8, SD=0.5\)), the Audio Volume category allowed participants to isolate and prioritize sounds. In Scene~2, for example, most participants adjusted the cats’ meows to identify differences. P3-VI described this as mirroring real-world strategies: \emph{``try to isolate one thing at a time and listen.''} P4-LP suggested volume control could help prioritize critical in-game alerts. 

\paragraph{Color.}  
Frequent and successful (\(M=4.6, SD=0.5\)), Color category prompts supported object recognition and contrast adjustments. For low-vision participants, recoloring enhanced readability and visibility (\eg{}P5-VI applied black-on-white contrast; P6-VI recolored objects to \emph{``see them and move them and understand what's there even better''}). For blind participants, relevance varied: P7-B noted limited utility without color knowledge, whereas P8-B, who had prior vision, used color references to mentally reconstruct scenes. Regarding potential improvements, some participants (N=2) suggested that the system should better recognize an object's background color to select an appropriate high-contrast color.

\paragraph{Object Size.}  
Moderately rated (\(M=4.1, SD=0.8\)), resizing aided low-vision players. P6-VI explained that enlarging objects made them easier to recognize and manipulate. For others, the numeric size descriptions drew mixed reactions. P7-B valued precise measurements in meters or feet, since these remain consistent regardless of viewing distance: \emph{``as you move away from an object, it visually appears smaller. If everything is in meters or feet... then [the description] is not affected by scale.''} In contrast, P8-B found numeric values difficult to interpret without visual reference. Some participants proposed alternatives: P2-VI suggested zooming in on objects rather than resizing them, while P5-VI highlighted the importance of knowing size limits (\eg{}maximum enlargement). Others, like P4-LP, considered size largely aesthetic and of limited accessibility value.

\paragraph{Scene Brightness.}  
Less frequently used and variably rated (\(M=3.6, SD=1.7\)), brightness adjustments helped some low-vision participants by enhancing contrast (P6-VI) or highlighting local areas (P5-VI). For blind participants, value was limited. For example, P4-LP dismissed it as \emph{``cosmetic.''} Usefulness thus depended strongly on visual ability.

\paragraph{Audio Pitch.}  
Lowest rated (\(M=2.8, SD=1.7\)), Audio Pitch adjustments were less useful overall. Some participants envisioned benefits for distinguishing status (\eg{}damage to a game character) or orientation (P7-B: \emph{``using pitch for positional information is a really good idea... for example, gets higher when you get closer to it''}), but most preferred volume as a more salient cue. {Participant comments suggested potential for system-level pitch cues for important events or orientation in future designs, but this category had limited value as a user-driven adjustment in our scenarios.}

\subsubsection{Emerging Categories}

\paragraph{Scene and Semantic Descriptions.}  
Although not predefined, these categories were frequently used, revealing additional details preferred by participants. Scene descriptions offered general overviews, while semantic descriptions probed higher-level qualities (\eg{}the mood of a cat or the content of a laptop screen). Participants valued detail but noted issues of verbosity: P4-LP suggested limiting output to nearby objects or allowing control over granularity. P7-B emphasized context-aware tailoring: \emph{``Maybe all I need to know is... I don't need to know that it's a Toshiba.''} Out-of-scope acknowledgments were common, as users often sought knowledge beyond system capabilities. Suggested improvements included adding shape (P1-LP), clearance (P7-B), sound characteristics (P2-VI, P8-B), and hazard indicators (P7-B).

\paragraph{Functionality and Creation/Deletion.}  
Although infrequent, these categories revealed unmet needs. Participants wanted to interact with objects functionally (\eg{}``sit on the bench'' rather than just move to its coordinates) and to add or remove objects. As P4-LP pointed out, without supporting object functionality interaction, the objects felt like ``dry artifacts'' instead of interactive video game objects. Success rates were low, especially for Creation/Deletion, reflecting system limitations rather than lack of interest. However, these attempts highlight aspiration for richer interactivity beyond current system capabilities.

\subsubsection{Category Usage Comparison Between Scene~2 and Scene~3}
\begin{figure*}
    \centering
    \includegraphics[
    width=1\linewidth,
    alt={The image shows the prompt category occurrence frequency by scene. On the x axis are the categories, and the y axis is the percentage of prompt categories. The graph shows percentage values for both Scene 2 and Scene 3.}
    ]{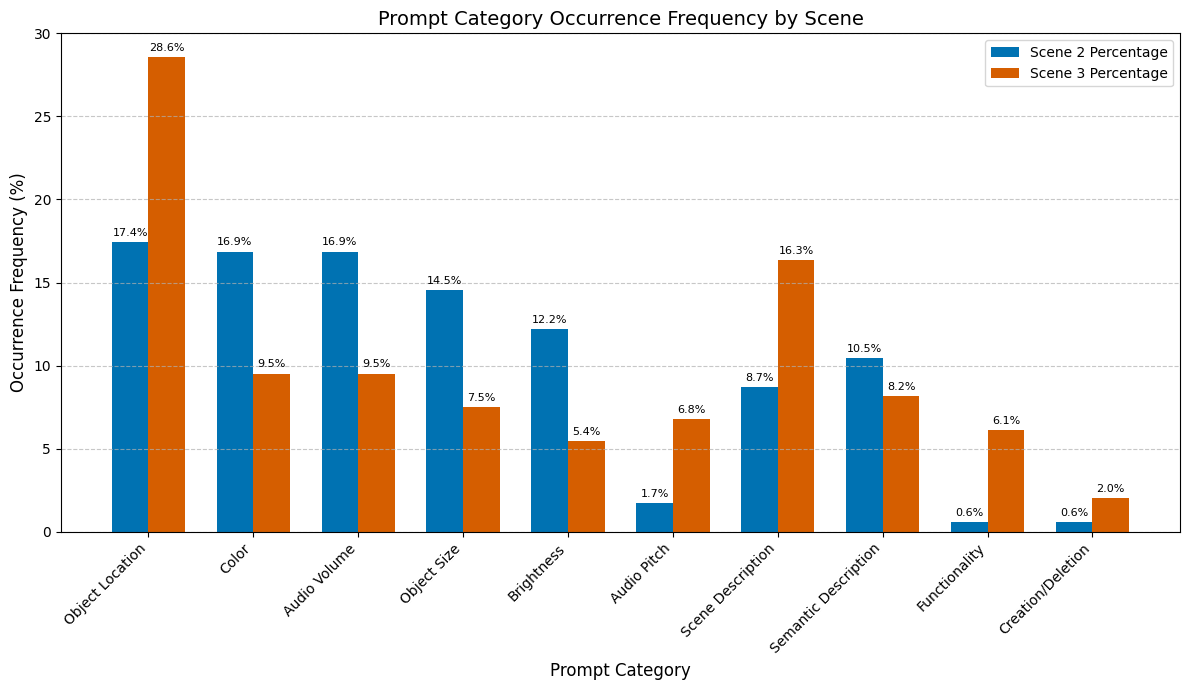}
    \caption{{Occurrence frequency of each category within the categorizable prompts (excluding \emph{Compound} and \emph{Other}) in each scene (blue = Scene~2 percentage, orange = Scene~3 percentage).}}
    \label{fig:scene-categories}
\end{figure*}

{Scene~2 asked participants to complete specific tasks aligned with the six guiding categories, whereas Scene~3 encouraged open-ended exploration. \Cref{fig:scene-categories} shows the percentage of prompts in each category for the two scenes (excluding \emph{Compound} and \emph{Other} prompts). Five categories showed notably higher relative usage in Scene~3 than in Scene~2: Object Location, Scene Description, Audio Pitch, Functionality, and Creation/Deletion. These differences suggest two broad exploration patterns in the open-ended scene: \emph{spatial exploration} (Object Location, Scene Description) and \emph{novel feature exploration} (Audio Pitch, Functionality, Creation/Deletion). Together, they highlight how participants appropriated \system{} differently when not guided by predefined tasks.}

\subsection{Emergent User Strategies and Goals}
\label{ssec:prompt-functionalities}

Across both the task-oriented scene (Scene~2) and the open-ended scene (Scene~3), participants used prompts serving different goals. Our coding produced eight user goal codes, grouped into three categories: \emph{exploration}, \emph{execution}, and \emph{verification} (see \cref{tab:functionalityCode} for details and examples).

\paragraph{Exploration.}  
Exploration was the entry point for nearly all tasks. It included US (understand scene), SS (search in scene), and QI (questions about specific items). In Scene~2, 33 of 48 tasks (68.8\%) began with exploration, and in Scene~3, all eight participants opened with such prompts. P5-VI explained: \emph{``You have to first ask what is an object, then know what it is, then increase size... and be more specific.''} Among the 336 total prompts, QI was most frequent (93), followed by SS (48) and US (36). These patterns highlight the central role of exploration in helping BLV users build an initial mental model of the environment. {The success rate for US prompts (86.1\%) was higher than for SS (75.0\%) and QI (73.1\%), suggesting that strategies starting with broad, whole-scene queries were more reliably supported than strategies that immediately targeted specific objects.}

\paragraph{Execution.}  
After establishing orientation, participants used execution prompts to act on the scene. Codes included EK (external knowledge), EM (explicit modification), PM (proactive modification), and CM (creative modification). EK was rare (7 prompts) but notable, as it showed users using the LLM to probe for accessibility knowledge beyond the scene itself, such as  \emph{``If I had a camera, would the light sources be bright enough to take a clear picture?''} P4-LP highlighted its promise: \emph{``you could have the game intuit its own design and say how might this be more accessible in a non-visual way?''} {EK prompts also demonstrated a relatively high success rate (85.7\%), underscoring that this strategy worked well when participants treated the LLM as an accessibility advisor.} 

EM and PM were common, aligning with task completion and proactive adjustments. Interestingly, CM appeared in 34 prompts, with participants using the system creatively, such as \emph{``Create a door in front of me''} (P2-VI) or \emph{``add a canopy and luxury padding to the bench''} (P7-B). For some, creative use became dominant — P2-VI devoted 10 of her 17 free-exploration prompts to CM. {Creative modifications were sometimes successful but also more likely to expose limitations (\eg{}unsupported functionality or implausible object behavior), indicating that CM is a powerful but less predictable strategy compared to EM and PM.} These findings suggest that beyond accessibility, participants viewed the system as a potential authoring tool.

\paragraph{Verification.}  
Verification prompts (V) were used to confirm whether modifications occurred as intended. They appeared 37 times overall, primarily in Scene~2 where tasks had specific targets. In Scene~3, use varied: some participants (N=5) regularly checked outcomes, while others (N=3) did not verify at all. P8-B issued six verification prompts across twelve modifications, explaining: \emph{``software applications or technology is not perfect, and that you just [have to] understand that there will be glitches.''} This reflects differing trust strategies: some relied on system feedback directly, while others double-checked modifications through dialogue.

\begin{table*}
  \caption{User goal codes across three groups.}
  \label{tab:functionalityCode}
  \begin{tabular}{ll p{5cm} p{6.3cm}}
    \toprule
    Group & Code & Description & Example \\
    \midrule
    \texttt Exploration & US & Understand whole scene & ``What is this scene like?'' \\
    \cellcolor[HTML]{dfdfdf}{Exploration} & \cellcolor[HTML]{dfdfdf}{SS} & \cellcolor[HTML]{dfdfdf}{Search within scene} & \cellcolor[HTML]{dfdfdf}{``Is there a white cat?'', ``Where is the loudest cat?''} \\
    \texttt Exploration & QI & Question about specific item(s) & ``How big is the conference table?'' \\
    \cellcolor[HTML]{dfdfdf}{Execution} & \cellcolor[HTML]{dfdfdf}{EK} & \cellcolor[HTML]{dfdfdf}{External knowledge} & \cellcolor[HTML]{dfdfdf}{``What color is useful for low-vision people?''} \\
    \texttt Execution & EM & Explicit modification (task-driven) & ``Make the bench bigger.'' \\
    \cellcolor[HTML]{dfdfdf}{Execution} & \cellcolor[HTML]{dfdfdf}{PM} & \cellcolor[HTML]{dfdfdf}{Proactive modification (user-driven)} & \cellcolor[HTML]{dfdfdf}{``Mute the other cats.''} \\
    \texttt Execution & CM & Creative modification & ``Make the chairs float in the air.'' \\
    \cellcolor[HTML]{dfdfdf}{Verification} & \cellcolor[HTML]{dfdfdf}{V} & \cellcolor[HTML]{dfdfdf}{Verify changes} & \cellcolor[HTML]{dfdfdf}{``What's the color of the chair now?''} \\
    \bottomrule
  \end{tabular}
\end{table*}



\section{Preliminary Evaluation with Unity Developers}
\label{sec:prelim-dev-study}

\subsection{Method}
\label{ssec:dev-study-method}

{To understand how \system{} fits into real development workflows and to evaluate its scalability from a developer perspective, we conducted a preliminary usability study with six Unity developers (three men, three women). Participants were recruited through forum posts and snowball sampling. Their Unity experience ranged from 0.5 to 6 years ($M{=}1.8$, $SD{=}1.9$). Two participants had prior experience with accessibility toolkits (\eg{}Meta XR accessibility tools), and two had used LLMs in game development.}

{Researchers first demonstrated how to integrate \system{} into a simple example Unity scene (five objects including two 3D objects, a sound source, a light source, and a text object) following the workflow described in \cref{sssec:SceneConstruction}. Participants then repeated this integration themselves in the example scene. Next, they incorporated \system{} into one of their own scenes, either self-created or selected from online resources. Written instructions accompanied the implementation steps, and participants were not given time constraints.}

{We recorded (1) the time required to achieve a working integration in the example scene and (2) the time needed to apply a functioning version of \system{} to a personal scene. Time was measured from the start of implementation to the first successful runtime test of an LLM prompt. Although we did not explicitly prompt developers to think aloud, several spontaneously verbalized thoughts about the system during the process. Because some personal scenes were large, we also collected estimates of how many objects would require tagging for full coverage.}

{After implementation, participants completed a questionnaire (5-point Likert-scale items on learnability, usability, perceived accessibility improvement, and intent for future use), followed by a semi-structured interview to capture experiences, preferences, and improvement suggestions. Sessions were conducted in person, lasted approximately one hour, and participants received \$30 compensation. Quantitative ratings were summarized using descriptive statistics. Two coders performed thematic analysis~\cite{guest_applied_2012}, producing 12 codes grouped under four themes—initial learning, developer interaction, system behavior, and perceived accessibility impact (see supplementary materials for full code book).}

\subsection{Findings}

{Developers spent an average of 265.8 seconds ($SD{=}92.6$) integrating \system{} into the example scene and 331.8 seconds ($SD{=}63.2$) applying a working version to their own scenes. Estimated tagging effort averaged 26.7 objects per scene ($SD{=}16.7$), though estimates varied with scene complexity and the tagging strategies developers preferred.}

{\textbf{Initial Learning and Setup.} Developers described the setup process as easy and intuitive. Learnability received a mean rating of 4.7 ($SD{=}0.5$). All six participants commented on the ease of initial system setup. Three appreciated the drag-and-drop design, while two noted that such simplicity could particularly help novices who may feel overwhelmed by other accessibility tools. A recurring challenge involved unclear variable names (\eg{}using \texttt{isMeta} to denote non-physical objects like sunlight); four participants recommended more transparent naming conventions.}

{\textbf{Developer Interaction and Tagging Strategies.} Usability ratings remained positive when applying \system{} to developers' own scenes ($M{=}4.3$, $SD{=}0.5$). Tagging strategies varied: some developers preferred tagging only interactable objects, while others advocated tagging everything visible for parity with sighted access. D4 noted that tagging hidden or narrative-sensitive elements (\eg{}easter eggs) might unintentionally spoil gameplay, whereas D5 suggested including a tagging field to mark the importance of each object. D6 argued that “a sighted person can see all the objects, so to provide an equivalent experience, we should tag all of them.”}

{\textbf{System Behavior and Desired Automation.} Developers suggested automating several aspects of the workflow, including: generating unique object names (D2), enabling batch tagging (D3), collecting renderers from nested objects (D4), and generating text descriptions from models or images (D6). D2 also suggested an interface for customizing model and prompt settings, though this might increase the learning curve.}

{\textbf{Perceived Accessibility Impact and Adoption.} Developers rated \system{}'s potential to improve accessibility highly ($M{=}4.7$, $SD{=}0.5$) and said they would use it in future projects ($M$=4.7, $S$=0.5). Five participants felt \system{} could meaningfully enhance BLV user experience. Interestingly, D3 described the system as a connector between developers and disabled users: developers supply descriptions and structural intent, while the LLM tailors them to user needs.}

{However, three participants raised concerns about LLM-generated errors, echoing concerns from prior work in runtime generation~\cite{jennings_whats_2024} and feedback from BLV participants. For future improvements, participants suggested features such as event-history tracking (D6), improved input/output methods (N{=}4), and using \system{} not only for modifying scenes but also for authoring accessible scenes (N{=}2).}



\section{Discussion}
\label{sec:discussion}

Generative AI introduces new programming paradigms and conversational forms of human-computer interaction. \system{} illustrates how these capabilities can extend accessibility in 3D environments by allowing BLV users to query and adapt scenes in natural language. Our findings show that participants found the system usable and intuitive, valuing its flexibility and tolerance of ambiguous input. At the same time, accuracy limitations surfaced, raising questions of trust and verification—findings consistent with known limitations of LLMs for accessibility \cite{glazko_autoethnographic_2023,adnin_i_2024}.

{Throughout the study, participants appropriated \system{} in different ways across Scenes~2 and 3, highlighting both task-driven use and open-ended exploration. These behaviors, together with the error profile we observed, inform how RAVEN should be positioned relative to existing accessibility tools and how future systems might provide safe, scalable support.} Building on these results, we discuss opportunities for safeguarding and verification, the potential role of LLMs as accessibility experts, ways to reduce developer effort through metadata automation, and broader implications for conversational programming. We conclude with limitations and directions for future work.

\subsection{Towards Safe and Trustworthy On-Request Generative Access}
Open-ended prompting enabled BLV participants to make diverse runtime modifications, but its open-endedness also created risks. Prior work in runtime behavior generation has noted developer concerns about ``game-breaking'' mechanics, such as deleting important objects or blocking pathways~\cite{jennings_whats_2024}. Our findings echo these risks: four emergent categories of prompts went beyond expected usage, often resulting in hallucinations or failures. Compounded by the difficulty BLV participants faced in verifying modifications, these results underscore the need for both automated and user-driven safeguards.

\textbf{Accessibility-Focused Guardrails.} Prior work has proposed multi-layer guardrails to constrain LLM behavior across domains~\cite{chennabasappa2025llamafirewallopensourceguardrail}. Extending this approach, accessibility-focused guardrails could ensure that generated modifications preserve both functionality and accessibility. Future work might adapt existing secure code-generation frameworks such as static analysis and constraint-based filters (\eg{}CodeShield~\cite{chennabasappa2025llamafirewallopensourceguardrail}) to restrict modifications that could regress accessibility, building on recent guidelines for accessible agentic interaction~\cite{fu_generative_2025,Zhang:2025:SaferAI,peng2025moraeproactivelypausingui}. 

\textbf{Automated Verification.}  
In our study, verification relied on the same LLM that generated modifications, risking false confirmations when errors occurred. Future systems could mitigate this by separating generation and verification roles, using multi-agent methods such as ensemble consensus~\cite{yang_one_2023,naik_probabilistic_2024} or debate-based techniques~\cite{smit_should_2024}. Multi-modal models could further enhance verification by cross-checking code execution against visual and auditory outputs from the scene~\cite{google_live_2025}. Though GPT-4o supports such multi-modal capabilities, we were unable to use them as these features were not available in the APIs at system development time.

\textbf{Human-in-the-Loop Verification.}  
Participants frequently issued follow-up prompts to confirm whether modifications had been applied, reflecting both natural interaction patterns and limited trust in the system. Prior work shows that LLMs can only audit accessibility in a limited capacity in well-standardized domains such as apps~\cite{Zhong:2025:ScreenAudit} and the web~\cite{Othman:2023:fostering}, underscoring the need for human oversight in open-ended contexts like 3D environments. Building on collaborative accessibility approaches in mainstream platforms (\eg{}Xbox’s controller assist), future systems could weave accessibility verification into mixed-ability collaborative gameplay. Such designs would preserve BLV user agency~\cite{Nair_navstick_2021,balasubramanian_enable_2023} while distributing responsibility for accuracy between users and systems, making accessibility verification an interdependent experience in virtual worlds~\cite{Branham:2015:CollaborativeAccessibility,Bennett:2018:Interdependence}. 

Taken together, guardrails, automated checks, and human oversight point to a layered verification strategy for ensuring both reliability and user trust in generative accessibility systems.

\subsection{LLMs as Accessibility Experts} 
Participants sometimes prompted the system as if it were an accessibility consultant, asking it to recommend or directly apply accessibility improvements (\eg{}``make the bench more visible to visually impaired individuals''). This highlights users’ expectation that LLMs can provide design knowledge beyond simple scene modifications. Prior work shows that LLMs can support accessibility tasks in domains such as web and app design~\cite{glazko_autoethnographic_2023,adnin_i_2024}, but also that they risk introducing ableist assumptions or biased recommendations. In 3D environments, this challenge is compounded by the need to scope recommendations to both user ability and scene context. Future systems could integrate guardrails that align suggestions with achievable in-system capabilities and contextual constraints, enabling LLMs to act as reliable accessibility advisors rather than overgeneralizing.

\subsection{Positioning RAVEN Among Accessibility Approaches}
\label{ssec:baseline-positioning}

{RAVEN occupies a distinct space among existing accessibility approaches for virtual 3D environments. Static toolkits such as {SeeingVR \cite{zhao_seeingvr_2019}} offer valuable enhancements for low-vision users but rely on developer-authored overlays and visual filters. These toolkits do not support runtime code generation, natural language prompting, or modification of audio and spatial structures. As a result, they are complementary to rather than directly comparable with \system{}, which focuses on enabling BLV users to interactively query and modify 3D environments in real time.}

{An alternative design question is whether large language models could simply “make the entire scene accessible” in a single transformation. Our findings suggest that such one-shot global modifications would not meet BLV users’ needs. Participants expressed diverse and sometimes conflicting accessibility preferences across the visual-ability spectrum—for example, some preferred brighter scenes while others needed dimmer ones, and some relied on color semantics while others valued purely spatial descriptions. Participants also engaged in iterative exploration, targeted adjustments, and verification, using conversational interaction to personalize accessibility in ways that a single global transformation could not capture. Importantly, accessibility needs emerged as highly individualized and context-dependent in our study, suggesting that adaptive, user-guided modification workflows are more appropriate than rigid, global transformations.}

{By enabling controlled, query-guided modifications grounded in embodied metadata and accessibility rules, \system{} supports accessible interaction as an iterative, user-directed process. This design preserves user agency, aligns modifications with contextual grounding, and avoids the brittleness of one-shot transformations, positioning \system{} as a complementary and necessary approach alongside existing static and LLM-based accessibility tools.}

\subsection{Reducing Developer Burden through Metadata Automation}  
{Our system required manual developer setup to populate the semantic scene graph with object and sound descriptions (\cref{sec:prelim-dev-study}). In our preliminary developer study, participants found the overall workflow learnable and usable, but expressed concern about scaling tagging and description efforts to larger scenes. Several requested automation to generate names, collect renderers from nested objects, and produce first-pass descriptions for both visual and audio elements.}  

{These requested capabilities can be supported in future integrations of \system{}'s authoring into mainstream developer environments. For example, Cap3D provides scalable 3D object captioning~\cite{luo_scalable_2023}, and ExCap3D supports expressive, variable-level descriptions~\cite{yeshwanth2025excap3dexpressive3dscene}. Audio language models (\eg{}Audio Flamingo~\cite{kong_audio_2024}) can generate captions for sound sources. Accessibility-focused plugins such as UI Accessibility Plugin~\cite{MetalpopGames_ui_2025} could integrate these intelligent capabilities to generate a first-pass layer of accessibility metadata for developers.}

Future systems could integrate such tools into development workflows (\eg{}Unity editor scripting~\cite{unitylearn_unity_2025}) to semi-automate tagging. Developers would still refine outputs to capture context-specific meaning (\eg{}labeling a “red vial” as a “health potion”), but automation could substantially reduce workload. By combining automated metadata generation with human curation, systems could better support detailed, context-aware accessibility queries without placing unrealistic demands on developers.

\subsection{Conversational Programming for Personalized Accessibility}
Beyond ensuring safety and scalability, our findings highlight the importance of personalization—how accessibility preferences differ across users and can be supported through conversational programming. \system{} is, to our knowledge, the first system to combine conversational interaction with runtime code generation for accessibility in 3D environments. Despite current limitations, participants’ engagement demonstrates the potential of conversational programming~\cite{Repenning:2011:ConversationalProgramming} to remediate accessibility barriers in ways that adapt to individual preferences, particularly where fixed guidelines fall short.

Our findings showed marked differences in how categories were valued depending on participants’ visual ability and history of vision. Low-vision participants benefited from color and brightness adjustments, while blind participants often dismissed these as cosmetic or irrelevant. Conversely, blind participants emphasized precise spatial orientation, whereas some low-vision participants preferred richer visual cues. By accommodating both, \system{} responds to the diverse and sometimes conflicting preferences of BLV users~\cite{szpiro_how_2016}, demonstrating how conversational programming can enable personalized accessibility. {Rather than treating “blind” and “low-vision” users as two homogeneous groups, our analysis reflects this continuum of abilities and preferences, which further motivates an adaptive, user-directed approach over static presets.}

Beyond 3D environments, similar approaches could extend to domains such as web browsing, education, or productivity tools. For example, extensible screen readers like NVDA could be augmented with LLM capabilities to generate custom add-ons from natural language instructions, tailoring themselves to individual needs. More broadly, the non-prescriptive and adaptive nature of conversational programming could support people with multiple disabilities or with fluctuating access needs, where requirements change across contexts and personal circumstances~\cite{mack_chronically_2022}.

\subsection{Limitations and Future Work}  
Our study represents an early exploration of conversational programming for accessibility in 3D environments. Several limitations qualify our findings and point to opportunities for future work. First, the system exhibited a non-trivial error rate, meaning it is not yet deployable in real-world applications revTwo{and may have negatively impacted user trust and usability during evaluation.} Future work should prioritize error prevention and robust verification methods to increase trust and reliability. Second, the system lacked an understanding of object affordances (\eg{}distinguishing the sitting surface of a bench from its coordinates), limiting functional interactions. Richer semantic modeling of object properties will be needed to support more realistic accessibility modifications. Third, our evaluation was conducted in a controlled, scenario-based setting with short-term tasks. {Although this is comparable to other accessibility evaluations, our sample size (eight BLV participants) limits the statistical power of between-group comparisons, which means our findings should be interpreted primarily as qualitative and exploratory.} Finally, longer-term deployments in commercial games or fast-paced contexts (\eg{}combat scenarios) are necessary to assess real-world viability.



\section{Conclusion}
\label{sec:conclusion}
\system{} explores a new frontier in accessible interaction—enabling blind and low-vision users to query and modify 3D virtual environments through natural language. By combining semantic scene understanding with real-time code generation, \system{} shifts accessibility from a static, developer-defined feature to an interactive, user-driven experience. Our evaluation with eight BLV participants demonstrated the promise of this approach: users found the system intuitive, flexible, and empowering. At the same time, limitations around reliability, error transparency, and emergent user needs point to the importance of future advances in guardrails, verification, and automated metadata. Such efforts will be critical to ensuring safe, trustworthy, and scalable deployment of generative accessibility systems.


\bibliographystyle{ACM-Reference-Format}




\appendix


\section{Appendix}
\subsection{Appendix 1: Prompt Constructor Instructions}
\label{ssec:appendix1}
\subsubsection{Accessibility Support Instructions:}
The following is a section about colors:
The HEX code represent the color. When asked about the color of an object, answer with natural language color instead of HEX code.
Red-green color blindness is a type of color vision deficiency that makes it difficult to distinguish between shades of red and green.
To make a scene more accessible for someone with red green color blindness, you should change the color palette. To make a palette for red green colorblind, avoid combining red and green. Also, make sure the new color created are not the same or similar as the other colors in the surroundings.
To highlight an object, you can use the GPT Indicator material.
To highlight an object without a Renderer, you create a transparent sphere with GPT Indicator material at its location for 5 seconds.
To select an object, you can create a transparent sphere with GPT Indicator material at its location for 5 seconds.
For simplifying material or texture of an object:
When asked to simplify material or texture of an object, create a new material that is closest to the object's original color and assign this new material to the object. If the original color is not provided, use the best guess given the object name.
To change the color of an object, first simplify the texture and then change the color.

The following is a section about object and text size:
When asked about size of an item, each unit is a meter. Answer how big an object is based on the size in meters.
When asked about how big is a text, answer the font size of the text.

The following is a section about spatial relationship between objects:
When ``me, I, my'' is referred, it means the player.
When asked about location of objects, answer the object's location relative to the player's location. 
When asked about the location of one object relative to another, respond by the distance calculated using euclidean distance between the center of the two objects. Be as precise as possible. Also answer how far the item is to the player in common sense, like "the object is close to you" or "the object is far away from you".

The following is a section about scene brightness:
To add a light to an area, create a Sphere game object in the area and add a point light to the sphere.
To make a light source brighter, adjust the intensity of the Light component.

The following is a section about audio sources:
To change the volume of a sound source, change the volume parameter on the AudioSource.
To change the pitch of a sound source, change the pitch parameter on the AudioSource.
To change the range of a sound source, change the max distance.

When asked to describe the scene or what are in the scene, describe it briefly, group similar objects together instead of listing all items.

\subsubsection{Error Prevention Instructions:}
If the request is general or incomplete, please ask follow-up questions for precise details and contexts, and leave the 'code' field null.
If the request says it's not working, please ask follow-up questions to clarify what's happening and suggest users to refine their request. If it's still not working, apologize to users and ask them to try another task.
If the request is out of your capability, tell users that the request is out of scope. The types of requests that cannot be achieved include: make zoom/magnifier, edge enhancement, color change on textured materials, object deletion.

\subsection{Appendix 2: Scene 1 Demo Example Prompts}
\label{ssec:appendix2}
\subsubsection{Color}
\begin{enumerate}
    \item ``What is the color of the cube?'' (query)
    \item ``What is the color of the sphere?'' (query)
    \item ``Make the color of the cube the same as the sphere.'' (modification)
    \item ``What is the color of the cube now?'' (verification)
\end{enumerate}

\subsubsection{Object Size}
\begin{enumerate}
    \item ``What is the size of the speaker 1?'' (query)
    \item ``Can you make it smaller?'' (modification)
    \item ``Will it fit into my hand?'' (verification)
\end{enumerate}

\subsubsection{Object Location}
\begin{enumerate}
    \item ``Grab one of the speakers onto my hand.'' (modification)
    *Move around and hear that speaker one is following you*
    \item ``What am I grabbing?'' (verification)
\end{enumerate}

\subsubsection{Scene Brightness}
\begin{enumerate}
    \item ``How bright is the scene?'' (query)
    \item ``Make the sunlight brighter.'' (modification)
    \item ``How bright is the scene now?'' (verification)
\end{enumerate}

\subsubsection{Audio Volume}
\begin{enumerate}
    \item ``Mute all speakers'' (modification)
    \item ``Unmute speaker one'' (modification)
    \item ``Move speaker one much closer to me'' (modification)
    \item ``Make the speaker one sound much louder'' (modification)
\end{enumerate}

\subsubsection{Audio Pitch}
\begin{enumerate}
    \item ``Make the pitch of speaker one higher.'' (modification)
    \item ``Is the pitch of speaker 1 higher than speaker 2 now?'' (verification)
\end{enumerate}


\end{document}